\documentclass[a4paper]{article}

\title{The Model Counting Competition 2020%
  \footnote{%
    The submitted solvers are available at
    \href{https://zenodo.org/record/4292581}{\nolinkurl{Zenodo:4292581}}~\cite{FichteHecherHamiti20b}.
    The benchmark instances for all tracks are available at
    \href{https://zenodo.org/record/3934427}{\nolinkurl{Zenodo:3934427}}~\cite{FichteHecher20a}.
    The contributed and collected instances are available
    at~\href{https://zenodo.org/record/4292168}{\nolinkurl{Zenodo:4292168}}~\cite{FichteHecherHamiti20}.
}
}

\author{Johannes K. Fichte \and Markus Hecher \and Florim Hamiti}

\date{%
  \today
}

\usepackage{geometry}
\usepackage{caption}
\usepackage{subcaption}
\usepackage{graphicx}
\usepackage{adjustbox}
\usepackage{microtype}
\usepackage{array}
\newcolumntype{H}{>{\setbox0=\hbox\bgroup}c<{\egroup}@{}}

\usepackage{fancyvrb}

\usepackage{booktabs}
\usepackage{amsmath,amsthm,thmtools,amssymb}
\newtheorem{example}{Example}

\declaretheoremstyle[
  headfont=\normalfont\scshape,
  numbered=unless unique,
  bodyfont=\normalfont,
  spaceabove=1em plus 0.75em minus 0.25em,
  prefoothook=\hfill\ensuremath{\blacksquare},%
  spacebelow=1em plus 0.75em minus 0.25em,
]{exmpstyle}

\declaretheorem[
  style=exmpstyle,
  title=Example,
  refname={example,examples},
  Refname={Example,Examples}
]{exmp}
\renewenvironment{example}{\begin{exmp}}{\end{exmp}}

\usepackage{xspace}
\newcommand{\SAT}{\texttt{SAT}\xspace}
\newcommand{\MC}{\texttt{MC}\xspace}
\newcommand{\WMC}{\texttt{WMC}\xspace}
\newcommand{\PMC}{\texttt{PMC}\xspace}
\newcommand{\ta}[1]{2^{#1}}
\newcommand{\eqdef}{\ensuremath{\,\mathrel{\mathop:}=}}
\newcommand{\SB}{\{}%
\newcommand{\SM}{\mid}%
\newcommand{\SE}{\}}%

\newcommand{\Card}[1]{\left|#1\right|}

\DeclareMathOperator{\wmc}{wmc}
\DeclareMathOperator{\pmc}{pmc}
\DeclareMathOperator{\var}{var}
\DeclareMathOperator{\lits}{lits}
\DeclareMathOperator{\Mod}{Mod}

\newcommand{\problemFont}[1]{\textsc{#1}}
\newlength\problemlength
\newcommand\dproblem[3]{%
\begin{center}
\fbox{%
\begin{minipage}{.93	\linewidth}%
\begin{list}{}{\labelwidth\problemlength \labelsep.7em \rightmargin1.5em
\leftmargin\problemlength \advance\leftmargin by3em%
\parsep0ex \itemsep.2ex plus.1ex}
\item[{\sl Problem:\hfill}] {\problemFont{#1}}
\item[{\sl Input:  \hfill}] #2
\item[{\sl Task: \hfill}] #3
\end{list}
\end{minipage}
}
\end{center}
}

\usepackage{xcolor}
\newcommand{\solver}[1]{\texttt{#1}}
\newcommand{\ops}[1]{{``\mbox{#1}''}}

\DeclareMathOperator{\cntc}{\#\cdot}%

\newcommand{\complexityclass}[1]{\ensuremath{\mathsf{#1}}}
\newcommand{\NP}{\ensuremath{\complexityclass{NP}}}

\newcommand{\cP}{\ensuremath{\cntc\complexityclass{P}}}
\newcommand{\cNP}{\ensuremath{\cntc\NP}}

\usepackage{hyperref}

\begin{document}
\maketitle

\begin{abstract}
  Many computational problems in modern society account to
  probabilistic reasoning, statistics, and combinatorics. A variety of
  these real-world questions can be solved by representing the
  question in (Boolean) formulas and associating the number of models
  of the formula directly with the answer to the question. 
  Since there has been an increasing interest in practical problem
  solving for model counting over the last years, the \emph{Model
    Counting (MC) Competition} was conceived in fall 2019.
  The competition aims to foster applications, identify new
  challenging benchmarks, and to promote new solvers and improve
  established solvers for the model counting problem and versions
  thereof. We hope that the results can be a good indicator of the
  current feasibility of model counting and spark many new
  applications.
  In this paper, we report on details of the Model Counting
  Competition 2020, about carrying out the competition, and the 
  results.
  The competition encompassed three versions of the model counting
  problem, which we evaluated in separate tracks.
  The first track featured the \emph{model counting problem (MC)},
  which asks for the number of models of a given Boolean formula.
  On the second track, we challenged developers to submit programs
  that solve the \emph{weighted model counting problem (WMC)}.
  The last track was dedicated to \emph{projected model counting
    (PMC)}.
  In total, we received a surprising number of 9 solvers in 34
  versions from 8 groups.
\end{abstract}

\section{Introduction}

\paragraph{Applications}
Many computational questions in modern society account to
probabilistic reasoning, statistics, and combinatorics. Examples of
such questions are autonomy for safety-critical
tasks~\cite{Fremont19a}, identifying the reliability of energy
infrastructure~\cite{MeelEtAl17a}, interactions in
bioinformatics~\cite{LatourEtAl17a}, recognizing
spam~\cite{ManningRaghavanSchutze08a,SahamiDumaisHeckerman98a},
optimizing budgeting in viral marketing~\cite{LatourEtAl17a}, learning
preference distributions~\cite{ChoiBroeckDarwiche15a}, carrying out
patient case simulations~\cite{PourretNaimBruce08a}, or predicting
weather~\cite{AbramsonBrownEdwards96a}.

A variety of these real-world questions can be solved by
\emph{representing} the question in (Boolean)
formulas~\cite{DomshlakHoffmann07a,SangBeameKautz05a,XueChoiDarwiche12a}
and associating the number of models of the formula directly with the
answer to the question.
Since there has been an increasing interest in practical problem
solving for counting the number of models over the last years, the
\emph{Model Counting (MC) Competition} was conceived in October 2019
to deepen the relationship between latest theoretical and practical
development on implementations for the model counting problem and
their practical applications.

The competition aims at identifying new challenging benchmarks, at
promoting new solvers, at improving established solvers for the model
counting problem and versions thereof, and at facilitating the
exchange of ideas and combining methods.
While this is a ``competition'' and we challenge researchers and
developers, there are no monetary prizes. 
We hope that active participation, collaboration, and long term
improvements extend the feasibility of model counting in practice
and spark many new applications.

We follow a direction in the community of constraint solving and
mathematical problem solving, where already many competitions and
challenges have been organized such as on
ASP~\cite{DodaroRedlSchuller19a} (7 editions),
CSP~\cite{RousselLecoutre19a,SimonisKatsirelosStreeter09a,TackStuckey19a}
(19 editions), SAT~\cite{HeuleJarvisaloSuda19a} (20 editions),
SMT~\cite{HadareanHyvarinenNiemetz19a} (14 editions),
MaxSAT~\cite{BacchusJarvisaloMartins19a} (14 editions),
UAI~\cite{GogateEtAl16a} (6 editions), QBF~\cite{PulinaSeidlShukla19a}
(9 editions), and various problem domains such as DIMACS (12
editions)~\cite{dimacs} and PACE (5
editions)~\cite{DzulfikarFichteHecher19a}.

\newcommand{\seq}[0]{Seq\xspace}
\newcommand{\para}[0]{Par\xspace}
\newcommand{\dist}[0]{Dist\xspace}

\newcommand{\sources}{\nolinkurl{Sources}}
\newcommand{\binary}{\nolinkurl{Binary}}

\begin{table}[t!]
  \centering
\begin{tabular}{llcccrlc}
  \toprule\\
  Problem & Solver &  Execution & Technique & Target & Reference & Download & License \\
  \midrule
  \MC & %
    \solver{ApproxMC}             & \seq & AC &  & \cite{ChakrabortyEtAl14a} & \href{https://github.com/meelgroup/approxmc}{\sources} & MIT\\
  & \solver{bdd\_{}minisat\_{}all} & \seq & BDD  & & \cite{TodaSoh15a} & \href{http://www.sd.is.uec.ac.jp/toda/code/cnf2obdd.html}{\sources} & MIT \\
  & \solver{c2d}               & \seq & KC & d-DNNF & \cite{Darwiche04a} & \href{http://reasoning.cs.ucla.edu/c2d/download.php}{\binary} & na \\
  & \solver{Cachet} & \seq & CC & & \cite{SangEtAl04} & 
                                                   \href{https://www.cs.rochester.edu/u/kautz/Cachet/cachet-wmc-1-21.zip}{\sources} & zchaff \\ %
  & \solver{d4}                    & \seq & KC & d-DNNF & \cite{LagniezMarquis17a} & \href{http://www.cril.univ-artois.fr/KC/d4.html}{\binary} & na\\
  & \solver{DSHARP}                & \seq & KC & d-DNNF &  \cite{MuiseEtAl12a} & \href{https://github.com/QuMuLab/dsharp}{\sources} & GPL2\\
  & \solver{Ganak}                & \seq & CC &  &  \cite{SharmaEtAl19a} & \href{https://github.com/meelgroup/ganak}{\sources} & MIT\\
  & \solver{miniC2D}               & \seq & KC  & SDD & \cite{OztokDarwiche15a} & \href{http://reasoning.cs.ucla.edu/minic2d/}{\binary} & na \\
  & \solver{cnf2eadt}              & \seq  & KC  & EADT & \cite{KoricheLagniezMarquisThomas13a} &  \href{http://www.cril.univ-artois.fr/KC/eadt.html}{\binary} & na\\
  & \solver{sdd}                   & \seq & KC & SDD & \cite{Darwiche11a} & \href{http://reasoning.cs.ucla.edu/sdd/}{\binary} & na \\ %
  & \solver{sharpCDCL} & \seq & CC & &  na &  \href{https://github.com/conp-solutions/sharpCDCL}{\sources} & GPL2\\
  & \solver{sharpSAT} & \seq & CC & & \cite{Thurley06a} & \href{https://sites.google.com/site/marcthurley/sharpsat}{\sources} & GPL2\\
  & \solver{sts} & \seq & AC &  & \cite{ErmonGomesSelman12a} & \href{http://cs.stanford.edu/~ermon/code/STS.zip}{\sources} & MIT \\
  & \solver{countAntom} & \para & CC &  & \cite{BurchardSchubertBecker15a} & \href{https://projects.informatik.uni-freiburg.de/projects/countantom}{\sources} & MIT\\
  & \solver{dpdb} & \para & DP & PTW & \cite{FichteEtAl20} & \href{https://github.com/hmarkus/dp_on_dbs/tree/padl2020}{\sources} & GPL3\\
  & \solver{gpusat} & \para & DP & PTW/ITW & \cite{FichteHecherZisser19a}  & 
                                                                              \href{https://github.com/daajoe/GPUSAT}{\sources} & GPL3\\
  & \solver{dCountAntom} & \dist & CC &  & \cite{BurchardSchubertBecker16a} & \href{}{na} & na \\
  & \solver{DMC} & \dist & KC & d-DNNF & \cite{LagniezMarquisSzczepanski18a} & \href{http://www.cril.univ-artois.fr/KC/dmc.html}{\binary} & na\\
  \midrule
  \WMC &
  \solver{ADDMC} &  \seq & DP & PTW+ADD & \cite{DudekPhanVardi20} & \href{https://github.com/vardigroup/ADDMC}{\sources} & MIT\\
  & \solver{Cachet} & \seq & CC & & \cite{SangEtAl04} & 
                                                   \href{https://www.cs.rochester.edu/u/kautz/Cachet/cachet-wmc-1-21.zip}{\sources} & zchaff \\ %
  & \solver{c2d}               & \seq & KC & d-DNNF & \cite{Darwiche04a} & \href{http://reasoning.cs.ucla.edu/c2d/download.php}{\binary} & na \\
  & \solver{d4}                    & \seq & KC & d-DNNF & \cite{LagniezMarquis17a} & \href{http://www.cril.univ-artois.fr/KC/d4.html}{\binary} & na\\
  & \solver{DPMC} & \seq & DP & PTW+ADD+TN& \cite{DudekPhanVardi20} & \href{https://github.com/vardigroup/DPMC}{\sources} & MIT\\
  & \solver{Ganak}                & \seq & CC &  &  \cite{SharmaEtAl19a} & \href{https://github.com/meelgroup/ganak}{\sources} & MIT\\
  & \solver{gpusat} & \para & DP & PTW/ITW & \cite{FichteEtAl18c}  & 
                                                                              \href{https://github.com/daajoe/GPUSAT}{\sources} & GPL3\\
  & \solver{TensorOrder} & \para & DP & TN & \cite{DudekVardi20} & \href{https://github.com/vardigroup/TensorOrder}{\sources} & MIT\\
  & \solver{miniC2D}               & \seq & KC  & SDD & \cite{OztokDarwiche15a} & \href{http://reasoning.cs.ucla.edu/minic2d/}{\binary} & na \\
  & \solver{sts} & \seq & AC &  & \cite{ErmonGomesSelman12a} & \href{http://cs.stanford.edu/~ermon/code/STS.zip}{\sources} & MIT \\
  \midrule
  \PMC 
   &  \solver{ApproxMC}             & \seq & AC &  & \cite{ChakrabortyEtAl14a} & \href{https://github.com/meelgroup/approxmc}{\sources} & MIT \\
  & \solver{nesthdb} & \seq & DP & ATW & \cite{HecherThierWoltran20} & \href{https://github.com/hmarkus/dp_on_dbs/tree/nesthdb}{\sources} & GPL3 \\         
  & \solver{projMC} & \seq & KC & d-DNNF  & \cite{LagniezMarquis19a} & \href{http://www.cril.univ-artois.fr/KC/projmc.html}{\binary} & na \\
  & \solver{ganak} & \seq & CC & & \cite{SharmaEtAl19a}& \href{https://github.com/meelgroup/ganak}{\sources} & MIT \\
  \bottomrule
\end{tabular}
\caption{%
  Overview on available model counting tools by problem domain%
}
\label{tab:mc-solvers}
\end{table}

\paragraph{The Problems and their complexity}
Given a Boolean formula the \emph{model counting problem}, \MC for
short\footnote{The model counting problem is also known as \#SAT.},
asks to output the number of models of that formula.
If in addition each literal in the formula has an associated weight
and the weight of a model is the product of its weights and we are
interested in the sum of weights over all models, we speak about
\emph{weighted model counting}, \WMC for short\footnote{The problem of
  weighted model counting is often also referred to as weighted
  counting, sum-of-product, partition function, or probability of
  evidence.}.
Another interesting counting problem is projected model counting, \PMC
for short. There, we hide some variables and we count the models after
restricting them to a set~$P$ of projection variables.
While the task of deciding whether a Boolean formula has a model
(\SAT) is already known to be \NP-complete~\cite{Cook71,Levin73}, its
generalization to counting is believed to be even harder. Namely, \MC
is known to be \cP-complete~\cite{Roth96a} and by direct implications
from a result by Toda~\cite{Toda91a} any problem on the polynomial
hierarchy~\cite{Stockmeyer76,StockmeyerMeyer73} can be solved in
polynomial-time by a machine with access to an oracle that can output
the model count for a given formula.
While \WMC is of similar complexity, \PMC is even harder assuming
standard complexity theoretical assumptions, more precisely, \PMC is
complete for the class $\cNP$~\cite{DurandHermannKolaitis05}.

\paragraph{Existing Solvers}
Many state-of-the-art solvers rely on standard techniques from
\SAT-based solving~\cite{%
  GomesKautzSabharwalSelman08a,%
  SangEtAl04,%
  Thurley06a%
}, knowledge compilation~\cite{LagniezMarquis17a}, or approximate
solving~\cite{ChakrabortyEtAl14a,ChakrabortyMeelVardi16a} by means of
sampling using \SAT solvers, and a few solvers employ dynamic
programming.
Table~\ref{tab:mc-solvers} provides a brief overview on recent solvers
and the used techniques. We give abbreviations in a footnote below.%
\footnote{ The abbreviations have the following meanings. %
    For the problem execution: 
    \seq (sequential solving), %
    \para (parallel solving), and %
    \dist (distributed solving). %
    For the techniques: 
    AC (approximate counting), CC (component caching), DP (dynamic programming), 
    KC (knowledge compilation).
    For details: 
    ADD (algebraic decision diagrams),  BDD (binary decision diagrams), SDD (Sentential Decision Diagrams), 
    ATW (abstraction treewidth), PTW (primal treewidth), ITW (incidence treewidth), TN (tensor networks), 
    EADT (extended affine decision trees), and d-DNNF (deterministic decomposable negation normal form).
  } %
  The website
  \href{http://beyondnp.org/pages/solvers/}{\nolinkurl{beyondnp.org}}
  provides a good overview.
There are also preprocessors available such as 
\href{http://www.cril.univ-artois.fr/kc/bpe2.html}{B+E}~\cite{LagniezLoncaMarquis16a}
and
\href{http://www.cril.univ-artois.fr/kc/pmc.html}{pmc}~\cite{LagniezMarquis14a}.
Many solvers are highly competitive and solve various
instances. However, there has still not been a competition on the
topics related to model counting, which spawned our interest in 
organizing one.

\section{The Problems}
Before we state the considered problems, we briefly provide formal
notions from propositional logic.  For a comprehensive introduction
and more detailed information, we refer to other
sources~\cite{BiereHeuleMaarenWalsh09,KleineBuningLettman99}.  Let $U$
be a universe of propositional variables.
A \emph{literal} is a variable~$x$ or its negation~$\neg x$. 
We call~$x$ \emph{positive} literal and $\neg x$ \emph{negative}
literal.
A
\emph{clause} is a finite set of literals, interpreted as the
disjunction of these literals.  A (Boolean) \emph{formula} (in
conjunctive normal form) is a finite set of clauses, interpreted as
the conjunction of its clauses.
We let $\var(F)$ and $\lits(F)$ be the set of the variables and set of
literals, respectively, that occur in~$F$.
An \emph{assignment} is a mapping $\tau:X \rightarrow \{0,1\}$ defined
for a set~$X\subseteq U$ of variables.
For $x\in X$, we define $\tau(\neg x)=1 - \tau(x)$. By $\ta{X}$ we
denote the set of all assignments~$\tau:X \rightarrow \{0,1\}$.
The formula~$F$ \emph{under assignment~$\tau$} is the formula~$F_\tau$
obtained from~$F$ by (i)~removing all clauses~$c$ that contain a
literal set to~$1$ by $\tau$ and then (ii)~removing from the remaining
clauses all literals set to~$0$ by $\tau$. An assignment~$\tau$
\emph{satisfies} a given formula~$F$ if $F_\tau=\emptyset$.
For a satisfying assignment~$\tau$, we call the set~$M$ of variables
that are assigned to true by~$\tau$ a \emph{model} of~$F$,~i.e.,
$M = \SB x \SM x \in \tau^{-1}(1) \SE$.

\begin{figure}[b]
  \centering
    \resizebox{0.33\textwidth}{!}{%
      \includegraphics{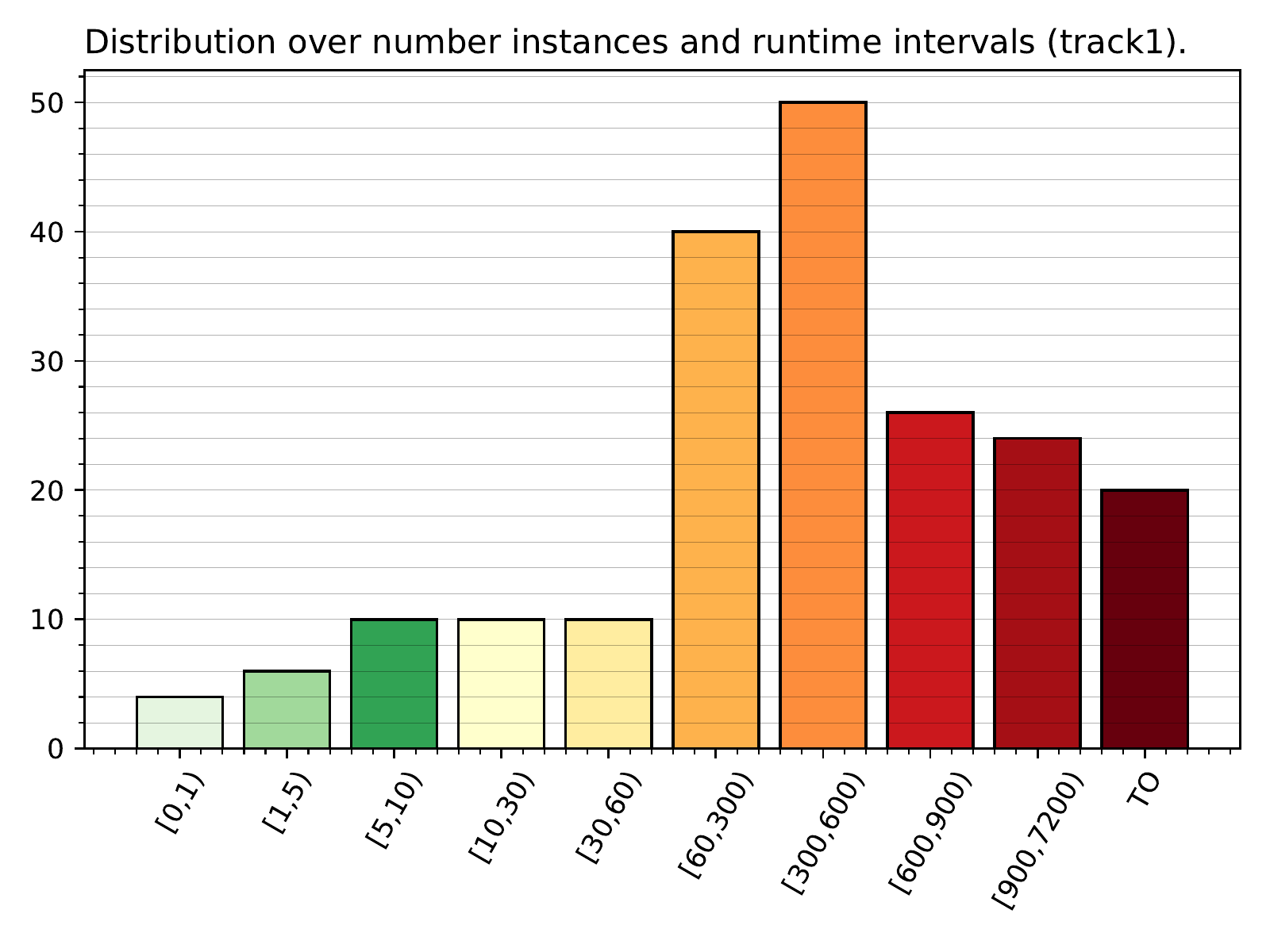}%
    }%
  \centering
    \resizebox{0.33\textwidth}{!}{%
      \includegraphics{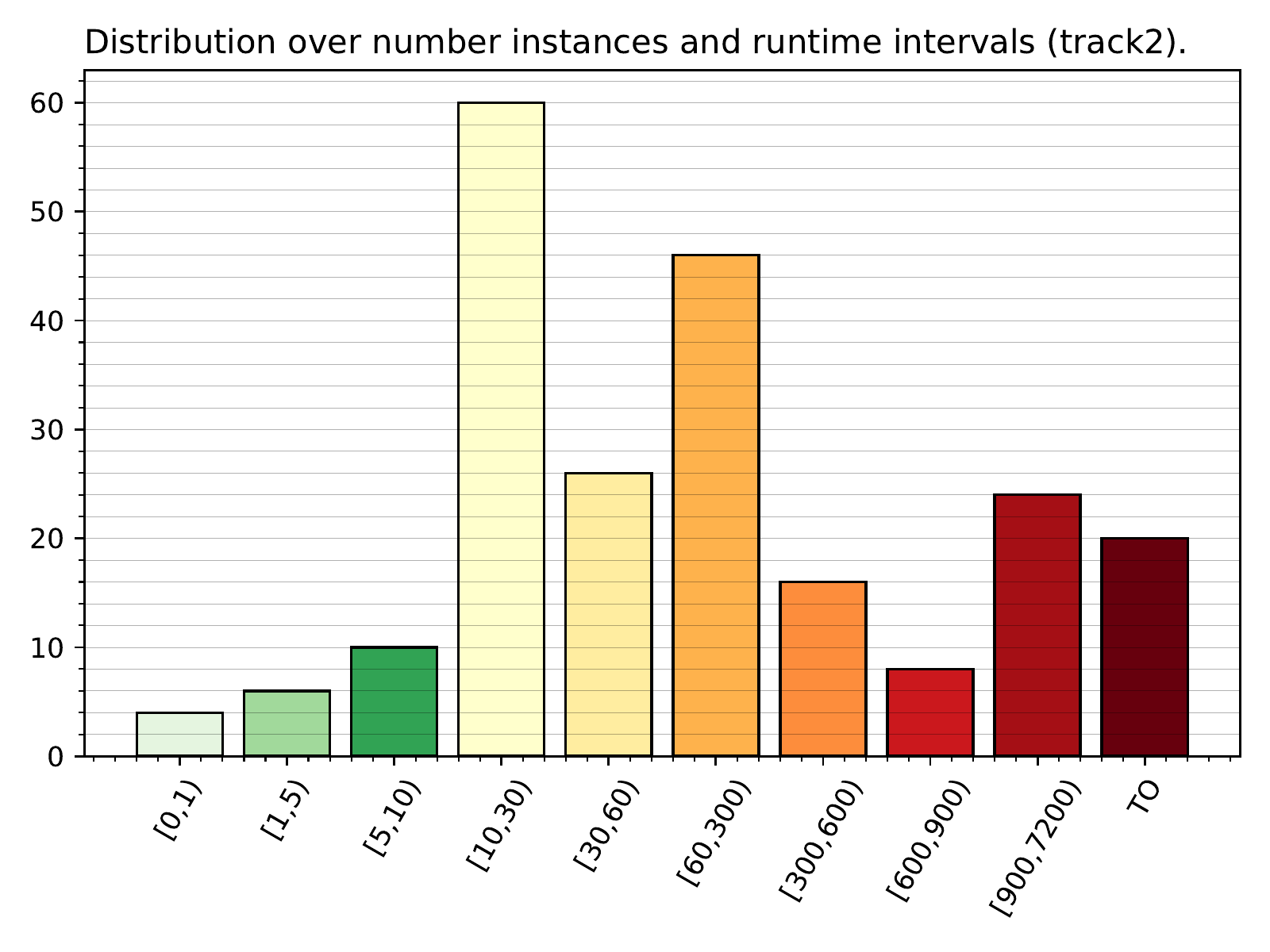}%
    }%
    \resizebox{0.33\textwidth}{!}{%
      \includegraphics{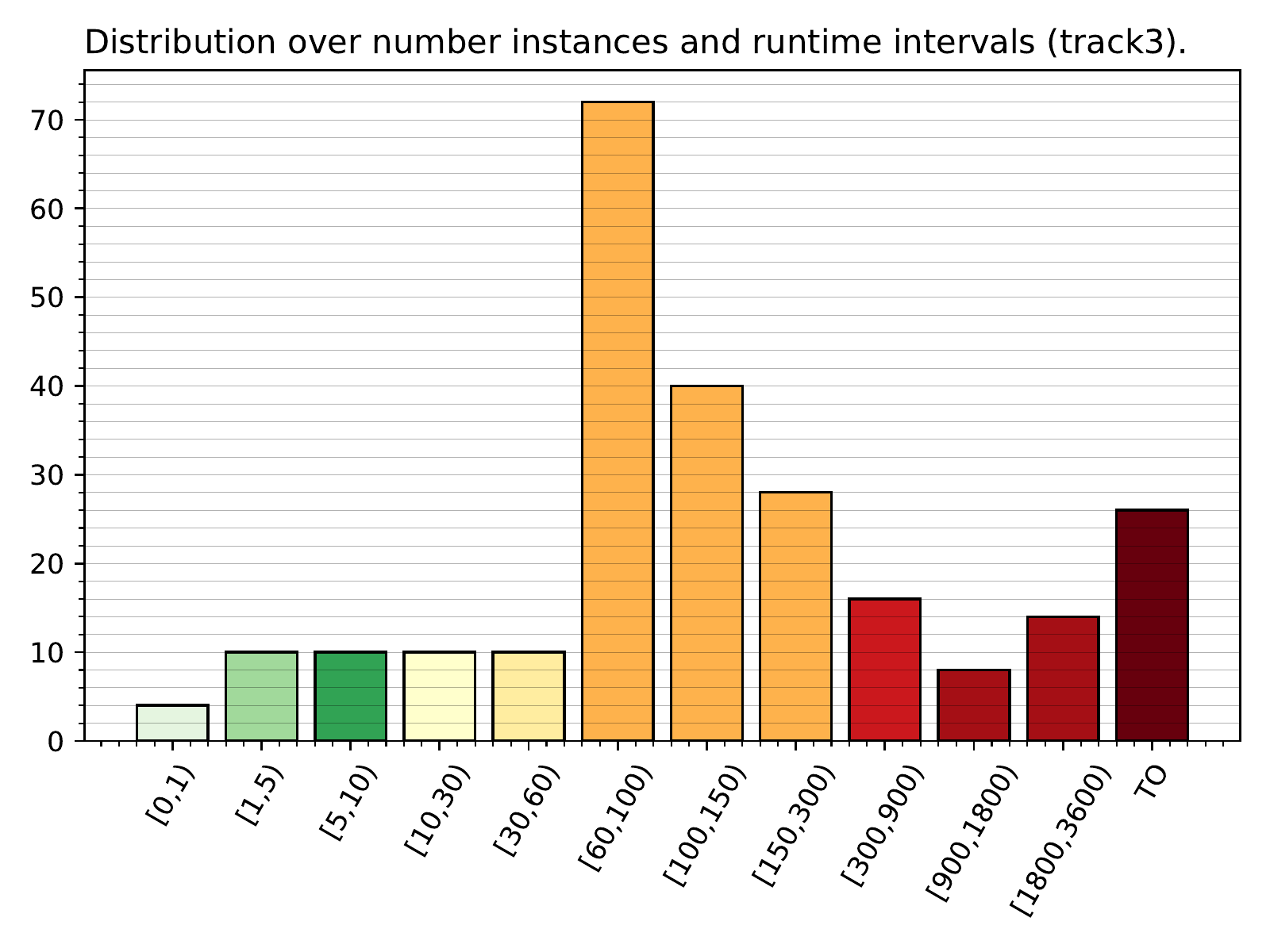}%
    }%
\caption{Distribution over the number of selected private instances for 
  Track~1 (MC),  Track~2 (WMC), and Track~3 (PMC) grouped by the used 
  hardness intervals. Colors indicate the ``practical'' hardness of the 
  instances (easy, medium, hard, very hard).
}
\label{fig:mc_runtime_distribution}
\end{figure}

\subsection{Model Counting (Track~1)}
The definitions from the previous section motivate the problem of
Track~1.

\dproblem{Model Counting (\MC)}%
{A Boolean formula~$F$ in conjunctive normal form.}%
{Output the number of models of the formula~$F$.}

\subsubsection*{Data Format}

The input format for providing a formula (\emph{.mcc2020\_cnf}) was
taken from the DIMACS-format for formulas in conjunctive normal
form~\cite{TrickChvatalCook93a}.\footnote{The DIMACS-input format is
  used in SAT competitions. For more details, we refer to an online
  resource at
  \url{http://www.satcompetition.org/2009/format-benchmarks2009.html}}
 We modified the problem description in the header to ``mc'' in order
to indicate that we aim for the model count.
More details on the format can be found in Appendix~\ref{sec:dataformat:mc}.

\subsubsection*{Instances}
In order to establish a suitable set of benchmark instances from
various areas, we posted an open call for benchmarks and collected
benchmarks from previously known sources.
Overall we received 1,220 instances from 6 groups in March
2020. Further, we took 1,619 instances from a benchmark collection
initiated by Daniel Fremont~\cite{FichteEtAl18b,Fremont20a}.
We did not check for duplicates.
We processed all instances by using the preprocessors
\href{http://www.cril.univ-artois.fr/kc/bpe2.html}{B+E~Apr2016}~\cite{LagniezLoncaMarquis16a}
and
\href{http://www.cril.univ-artois.fr/kc/pmc.html}{pmc~1.1}~\cite{LagniezMarquis14a},
separately. Then, we included the unpreprocessed, the preprocessed
instances (B+E), and the preprocessed instances (pmc).
For pmc, we used the documented options \ops{-vivification}
\ops{-eliminateLit} \ops{-litImplied} \ops{-iterate=10} \ops{-equiv}
\ops{-orGate} \ops{-affine}.
We started B+E with the option~\ops{-limSolver=1000}.
\newcommand{\bench}[1]{\texttt{#1}}
The contributors and origins of the instances are as follows:

\begin{enumerate}
\item 100 instances from \bench{VLSAT}
  (\href{cadp.inria.fr/resources/vlsat/}{\nolinkurl{https://cadp.inria.fr/resources/vlsat/}}),
  which are obtained by decomposing Petri nets into networks of
  automata contributed by Bouvier and
  Garavel~\cite{BouvierGaravel20a};
\item 303 instances from 5 domains including 
  \begin{enumerate}
  \item 29 instances from \bench{Bayes} (\href{https://www.cril.univ- artois.fr/KC/benchmarks.html}{\nolinkurl{cril.univ-artois.fr}}),
  \item 32 instances from \bench{Blasted SMT} (\href{https://github.com/meelgroup/sampling-benchmarks}{\nolinkurl{github:meelgroup/sampling-benchmarks}}), 
  \item 135 instances from \bench{Functional synthesis} (\href{http://www.qbflib.org/qbfeval18.php}{\nolinkurl{qbflib.org}}), 
  \item 54 instances from \bench{Program synthesis}
    (\href{https://github.com/dfremont/counting-
      benchmarks/tree/master/benchmarks/projection/application}{\nolinkurl{github:dfremont/counting-benchmarks}}),
    and
  \item 53 instances from \bench{Quantitative information flow} (\href{https://github.com/dfremont/counting-benchmarks/tree/master/benchmarks/projection/application}{\nolinkurl{github:dfremont/counting-benchmarks}})
  \end{enumerate}
  contributed by Lai, Golia, and Meel~\cite{FichteHecherHamiti20};
\item 596 instances from \bench{Program Analysis}
  (\href{https://doi.org/10.5281/zenodo.4292167}{\nolinkurl{zenodo.4292167}}),
  which are generated by a symbolic execution bug finding tool called
  CAnalyze contributed by M\"ohle, Ge, and
  Biere~\cite{FichteHecherHamiti20};
\item 12 instances from \bench{RandomHard}
  (\href{https://doi.org/10.5281/zenodo.4292167}{\nolinkurl{zenodo.4292167}}),
  which are randomly generated benchmarks aiming for increasing
  difficultly with increasing its number of literals contributed by
  Spence~\cite{FichteHecherHamiti20};
\item 70 instances from \bench{ModelRB}
  (\href{https://doi.org/10.5281/zenodo.4292167}{\nolinkurl{zenodo.4292167}}),
  which are randomly generated instances that resemble many hard
  instances for exact phase transitions in CSP contributed by Yin,
  Wang, and Xu~\cite{FichteHecherHamiti20}; and
\item 139 instances from \bench{RandomGen}
  (\href{https://doi.org/10.5281/zenodo.4292167}{\nolinkurl{zenodo.4292167}}),
  which are randomly generated instances for testing solvers including
  instances on graph coloring, monotone CNFs, and 2CNFs contributed by
  Pehoushek~\cite{FichteHecherHamiti20};
\item 1,619 instances from \bench{dfremont}
  (\href{https://github.com/dfremont/counting-
    benchmarks/tree/master/benchmarks/projection/application}{\nolinkurl{github:dfremont/counting-benchmarks}}),
  which is a collection of various benchmark instances originating in
  multiple domains retrieved from the collection by
  Fremont~\cite{Fremont20a}.
\end{enumerate}

\begin{figure}[t]
  \begin{subfigure}{1\textwidth}
  \centering %
    \begin{tabular}{lrrrrrrrr}
      \toprule
      instances &      $n_{\min}$ &   $n_{\max}$ &   $n_\text{avg}$ & $n_\text{med}$ &    $m_{\min}$ &   $m_{\max}$ &   $m_\text{avg}$ & $m_\text{med}$ \\
      \midrule
      public  &      1.0 &  7.7M &   113k &  3.6k &    1.0 &  5.7M &  95.2k &  5.3k \\
      private &      78.0 &   205.0k &  42.6k &  3.6k &    78.0 &   539k &  40.5k &  5.6k \\
      all     &      1.0 &  7.7M &  77.8k &  3.7k &    1.0 &  5.7M &  67.8k &  5.4k \\
      \bottomrule
    \end{tabular}
    \caption{Overview on the number of variables and clauses for the
      public, private, and all instances. $n$ and $m$ represent the
      number of variables and clauses, respectively. max refers to the
      maximum; avg refers to the mean; med refers to the median.}
  \end{subfigure}
  \begin{subfigure}{0.49\textwidth}
    \centering
    \resizebox{1\textwidth}{!}{%
      \includegraphics{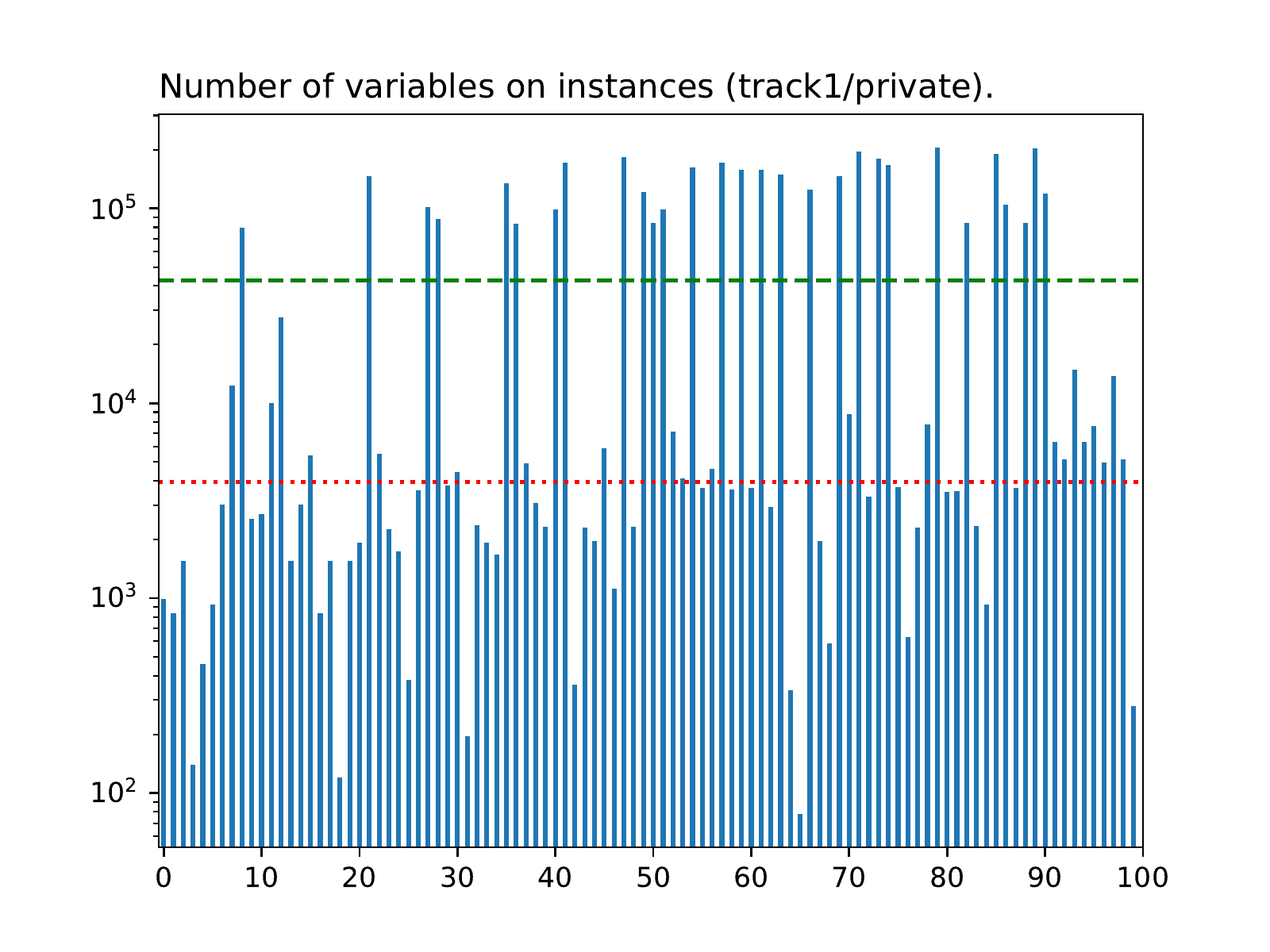}
    }%
  \caption{Overview on the number of variables of the private
    instances. The red dotted line indicates the median over number of
    variables and the green dashed line represents the mean over
    number of variables.}
  \end{subfigure}\hfill
  \begin{subfigure}{0.49\textwidth}
    \centering
    \resizebox{1\textwidth}{!}{%
      \includegraphics{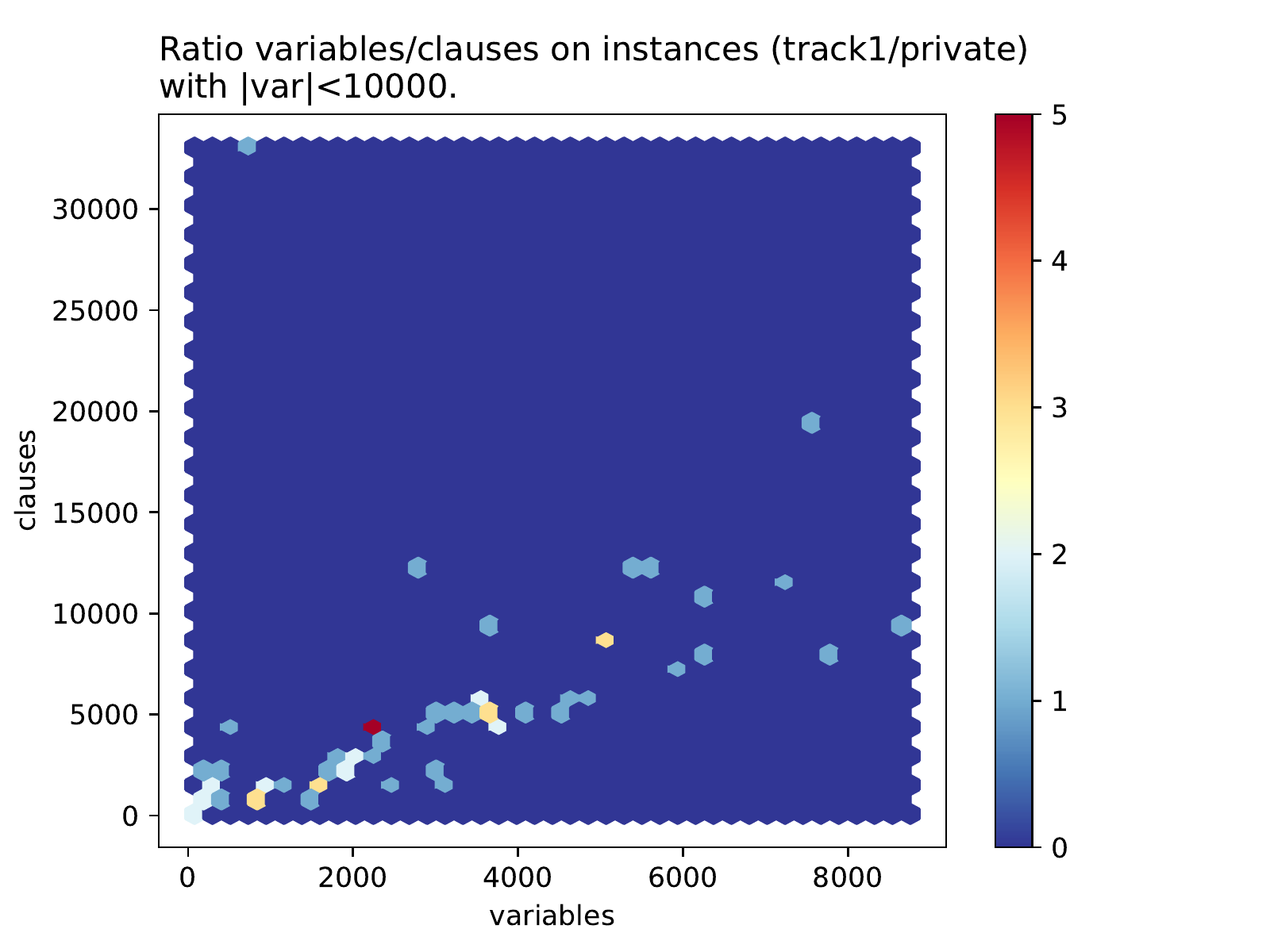}
    }%
    \caption{Ratio between the variables and clauses on the private
      instances when the number of variables are restricted to 10k.}
  \end{subfigure}
  \caption{Basic statistics on the selected instances for MC2020/Track1 (Model Counting).
  }
\label{fig:mc_stats}
\end{figure}

For the model counting competition, we are interested in challenging
instances of varying size that are still within reach for the
participants, but challenge them and require reasonable efforts to
score a high rank.
In order to get a naive classification of the ``practical hardness''
of the collected benchmark instances, we ran the solvers
\solver{Cachet}~\cite{SangEtAl04}, \solver{c2d}~\cite{Darwiche04a},
\solver{d4}~\cite{LagniezMarquis17a},
\solver{GANAK}~\cite{SharmaEtAl19a}, and
\solver{sharpSAT}~\cite{Thurley06a} on all instances with a timeout of
2 hours.
After obtaining initial runtime results, we assigned to each instance
a \emph{hardness category} (very-easy$\{1,2,3\}$, easy$\{1,2\}$, medium$\{1,2\}$, and
hard$\{1,2,3\}$).
From the classified instances, we chose 200 instances by sampling
uniformly at random among the distribution given in
Figure~\ref{fig:mc_runtime_distribution}.
In more detail, we selected very few instance with runtime below 1s
and overall picked 20 instances from category very-easy (runtime
within the interval $[0;10)$, given in seconds). We chose 20 instances from category
easy (runtime within the interval~$[10; 60)$, 90~instances from
category medium (runtime in the interval~$[60; 600)$), and 70
instances from category hard (runtime in the interval~$[$600;
7,200$)$). Among the hard instances are 20~instances for which we did
not obtain a solution within 7,200 seconds.
We numbered the instances from 1 to 200 with increasing hardness, 
selected the odd numbered instances as private and even numbered
instances as public instances. Figure~\ref{fig:mc_stats} shows
statistics on the resulting instances for Track~1.
The 100 public instances were disclosed at
\href{https://mccompetition.org/}{\nolinkurl{mccompetition.org}} in late April.
Both public and private instances are available for download at
\href{https://zenodo.org/record/3934427}{\nolinkurl{Zenodo:3934427}}~\cite{FichteHecher20a},
which also contains the mapping of the selected instances and the
original instance.

\begin{figure}[t]
  \begin{subfigure}{0.5\textwidth}
  \centering %
    \begin{tabular}{lHHrHrrrr}
      \toprule
      instances &      $n_{\min}$ &   $n_{\max}$ &   $n_\text{avg}$ & $n_\text{med}$ &    $m_{\min}$ &   $m_{\max}$ &   $m_\text{avg}$ & $m_\text{med}$ \\
      \midrule
      public  &       100 &  12.3k &  2.03k &    200 &     181 &  5.7k &  1.26k &    432 \\
      private &       100 &  12.3k &  1.47k &    200 &     216 &  12.8k &  1.19k &    398 \\
      all     &       100 &  12.3k &  1.75k &    200 &     181 &  12.8k &  1.22k &    418 \\
      \bottomrule
    \end{tabular}
    \caption{Overview on the number of variables and clauses for the
      public, private, and all instances. $n$ and $m$ represent the
      number of variables and clauses, respectively. max refers to the
      maximum; avg refers to the mean; med refers to the median.}
  \end{subfigure}
  \begin{subfigure}{0.49\textwidth}
    \centering
    \resizebox{1\textwidth}{!}{%
      \includegraphics{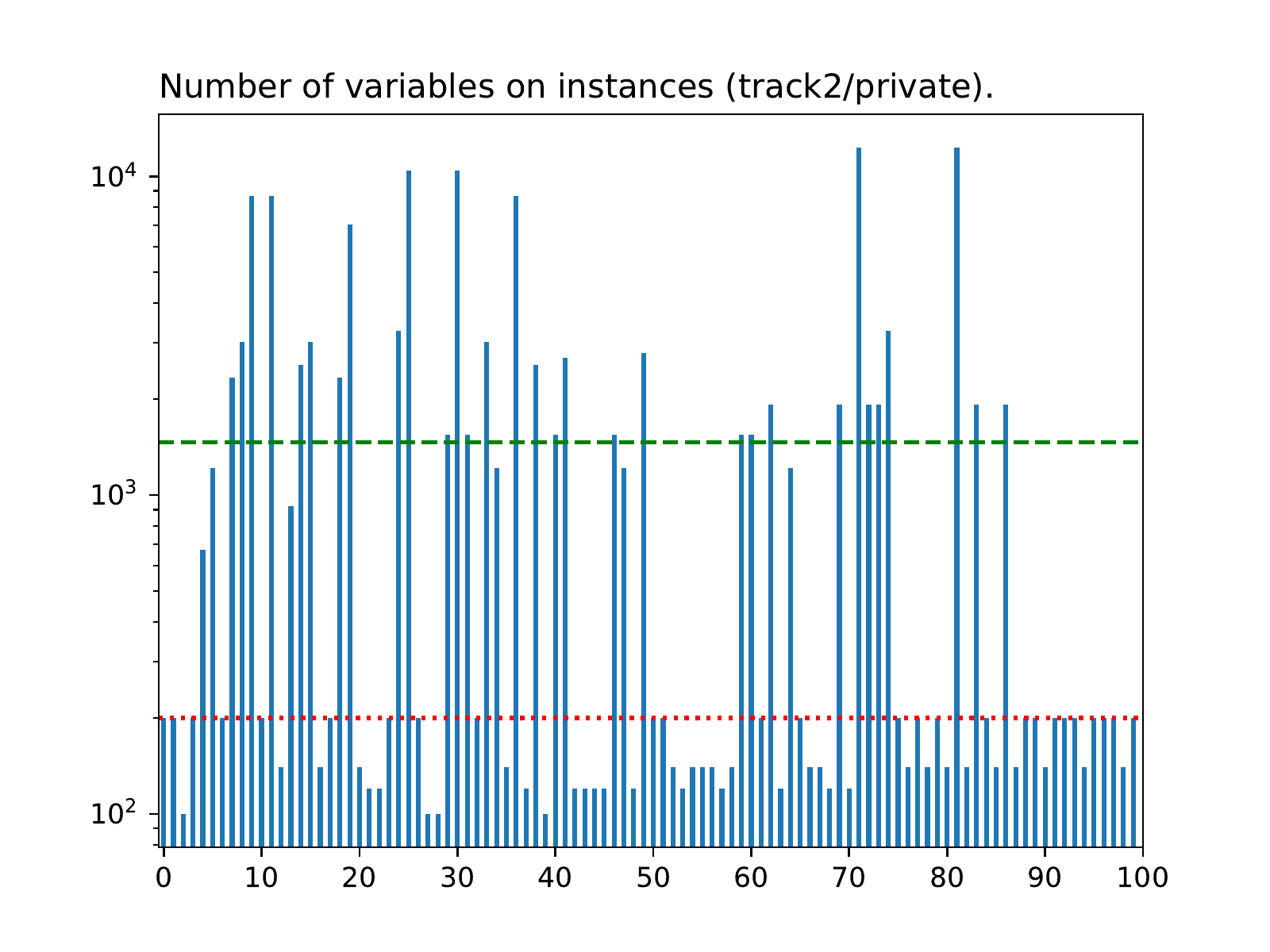}
    }%
  \caption{Overview on the number of variables of the private
    instances. The red dotted line indicates the median over number of
    variables and the green dashed line represents the mean over
    number of variables.}
  \end{subfigure}\hfill
\caption{Basic statistics on the selected instances for MC2020/Track2 (Weighted Model Counting).}
\label{fig:wmc_stats}
\end{figure}

\subsection{Weighted Model Counting (Track 2)}
Weighted model counting generalizes model counting as follows.
Let $F$ be a formula and assume that $\Mod(F)$ denotes the set of models
of~$F$.
We let a \emph{weight function}~$w$ be a function that maps each
literal in $F$ to a real between~$0$ and~$1$,~i.e.,
$w: \lits(F) \rightarrow [0,1]$.
While one often restricts  $w(v) + w(\neg v) = 1$, we explicitly allow
$0 \leq w(v) + w(\neg v) \leq 2$.
Then, for an assignment~$\tau$ to the variables in $F$, the
\emph{weight of the assignment~$\tau$} is the product over the weights
of its literals,~i.e.,
\[w (M) := \prod_{v\in \var(F) \cap M} w(v) \cdot \prod_{v \in \var(F)
    \setminus M} w(\neg v).\]%

\noindent The \emph{weighted model count} (wmc) of formula is the sum of weights
over all its models,~i.e.,%
\[\wmc(F,w) \eqdef \sum_{M \in \Mod(F)} w(M).\]

\dproblem{Weighted Model Counting (\WMC)\footnote{
    The problem is sometimes also called sum-of-products, weighted
    counting, partition function, or probability of evidence.
}}%
{A Boolean formula~$F$ in conjunctive normal form and a weight function~$w$.}%
{Output the weighted model count~$\wmc(F,w)$.}

\paragraph{Data Format}
The input format for providing a formula (\emph{.mcc2020\_wcnf}) was
taken from the DIMACS-format for formulas in conjunctive normal
form~\cite{TrickChvatalCook93a} and its modification in the solver
Cachet~\cite{SangEtAl04}.\footnote{The description for format that
  Cachet uses for weighted model counting is given in the file
  ``README.txt'' of the solver package, which is available online at
  \url{https://www.cs.rochester.edu/u/kautz/Cachet/cachet-wmc-1-21.zip}.} %
 We modified the problem description in the header to ``wmc'' in order
to indicate that we aim for the weighted model count.
In contrast to cachet, we define the weight function slightly
different.
Weights are given explicitly unless we assume that both the positive
and negative literal have weight~1,~i.e.,~$w(v)=w(\neg v) = 1$ for
variable~$v$.
If a weight is stated for a literal, then we assume both the weight
for the positive and negative literal are given.
For a literal~$\ell$, we provide weights as floating point numbers
between~$0 \leq w(\ell) \leq 1$.
More details on the format can be found in Appendix~\ref{sec:dataformat:wmc}.

\paragraph{Instances}%
For this track, we took publicly
available~\cite{FichteEtAl18b,Fremont20a} instances and modified some
weights. Overall the set consists of 1091~instances.
While one can apply the preprocessor pmc~\cite{LagniezMarquis14a} also
to \WMC\footnote{According to the documentation of pmc at
  \url{http://www.cril.univ-artois.fr/kc/pmc.html} the options
  \ops{-vivification} \ops{-eliminateLit} \ops{-litImplied}
  \ops{-iterate=10} preserve all models.}, we did not apply
preprocessing here. Further, we did not check for duplicates.
Similar to the previous track, we first estimated the ``practical
hardness'' of the instances. Therefore, we used the solvers
\solver{Cachet}~\cite{SangEtAl04},
\solver{d4}~\cite{LagniezMarquis17a}, and
\href{http://reasoning.cs.ucla.edu/minic2d/}{\solver{miniC2D}}~\cite{OztokDarwiche15a}
on all instances with a timeout of 2 hours.
From the classified instances, we chose 200 instances by sampling
uniformly at random among the distribution given in
Figure~\ref{fig:mc_runtime_distribution}.
We numbered the instances from 1 to 200 with increasing hardness, 
selected the odd numbered instances as private and even numbered
instances as public instances. Figure~\ref{fig:wmc_stats} shows
statistics on the resulting instances for Track~2.
The 100 public instances were disclosed at
\href{https://mccompetition.org/}{\nolinkurl{mccompetition.org}} in beginning of May.
Both public and private instances are available for download at
\href{https://zenodo.org/record/3934427}{\nolinkurl{Zenodo:3934427}}~\cite{FichteHecher20a}.
The links also refer to a document that contains the mapping of the
selected instances and the original instance.

\subsection{Projected Model Counting (Track 3)}
While the previous two tracks featured the model counting
problem and its weighted version, we might have situations during
modeling where we have to introduce auxiliary variables that are
important for the satisfiability of the formula, but they increase the
number of solutions and should not be counted.
So multiple solutions that include auxiliary variables count as just
one solution for us if we ignore the auxiliary variables.
However, if we are only interested to obtain the number of solutions
with respect to the variables of interest, we generalize the problem
to projected model counting as follows.
Therefore, let $F$ be a Boolean formula and $P \subseteq \var(F)$ be a
set of variables, called \emph{projection variables}.
We define the projected model count~$\pmc_P(F)$ of the formula by
\[\pmc(F,P) \eqdef\Card{\SB M \cap P \SM M \in \Mod(F) \SE}.\]
This gives then raise to the following problem:

\dproblem{Projected Model Counting Problem (\PMC)\footnote{Sometimes
    the problem is referred to as $\#\exists\SAT$ and was originally
    coined under the name \#NSAT, for ``nondeterministic SAT'' by
    Valiant~\cite{Valiant79b}.}}%
{A Boolean formula~$F$ in conjunctive normal form and a
  set~$P\subseteq \var(F)$ of \emph{projection variables}.}%
{Output the projected model count~$\pmc(F,P)$.}

\paragraph{Data Format}
The input format for providing a formula (\emph{.mcc2020\_pcnf}) was
taken from the DIMACS-format for formulas in conjunctive normal
form~\cite{TrickChvatalCook93a} and its modification as used in the
solver Ganak~\cite{SharmaEtAl19a}.
We modified the problem description in the header to ``pmc'' in order
to indicate that we aim for the projected model count.
Further, we indicate projection variables by a line starting with
``vp'' followed by the respective variables and terminated by a~0.
More details on the format can be found in Appendix~\ref{sec:dataformat:pmc}.

\begin{figure}[t]
  \begin{subfigure}{0.5\textwidth}
  \centering %
    \begin{tabular}{lHHHHrHrrrrrr}
      \toprule
      instances &      $n_{\min}$ &   $n_{\max}$ &   $n_\text{avg}$ & $n_\text{med}$ &    $m_{\min}$ &   $m_{\max}$ &   $m_\text{avg}$ & $m_\text{med}$ & $p_{\min}$ &   $p_{\max}$ &   $p_\text{avg}$ & $p_\text{med}$ \\
      \midrule
      public  &       168 &  869k &  216k &   331k &     385 &  1.56M &  355k &   586k &       28 &  726 &  138 &    100 \\
      private &       431 &  861k &  200k &   146k &   1.15k &  1.55M &  325k &   184k &       42 &  813 &  136 &    100 \\
      all     &       168 &  869k &  208k &   331k &     385 &  1.56M &  340k &   557k &       28 &  813 &  137 &    100 \\
      \bottomrule
    \end{tabular}
    \caption{Overview on the number of variables and clauses for the
      public, private, and all instances. $m$ and $p$ represent the
      number of clauses and projection variables, respectively. max refers to the
      maximum; avg refers to the mean; med refers to the median.}
  \end{subfigure}
  \begin{subfigure}{0.49\textwidth}
    \centering
    \resizebox{1\textwidth}{!}{%
      \includegraphics{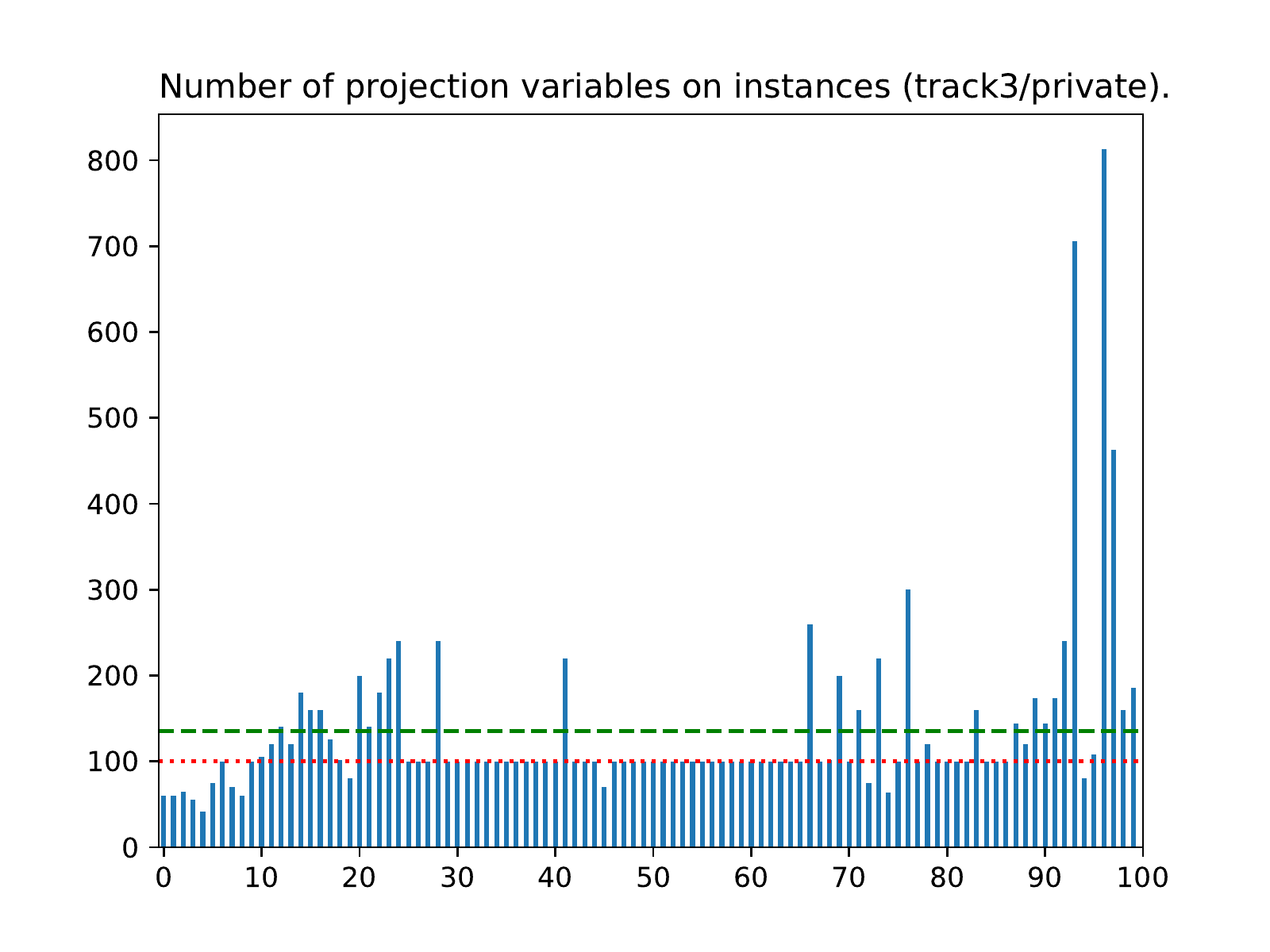}
    }%
  \caption{Overview on the number of projection variables of the private
    instances. The red dotted line indicates the median over number of
    variables and the green dashed line represents the mean over
    number of variables.}
  \end{subfigure}\hfill
  \caption{Basic statistics on the selected instances for
    MC2020/Track3 (Projected Model Counting).
  }
\label{fig:pmc_stats}
\end{figure}

\paragraph{Instances}%
Similar to the previous tracks, we posted an open call to submit new
benchmark sets. We received two submissions and included in addition the
publicly available instances~\cite{Fremont20a} instances. Overall the
set consists of 985~instances.
We applied the preprocessor pmc~\cite{LagniezMarquis14a} using the
options \ops{-vivification} \ops{-eliminateLit} \ops{-litImplied}
\ops{-iterate=10} to the sets \bench{dfremont} and \bench{2-QBF} and
included the resulting instances. Due to time constraints and runtime
limitations on the cluster, we were unable to apply the preprocessing
to the benchmark set \bench{Neural}.
We did not check for duplicates.
The contributors and origins of the instances are as follows:

\begin{enumerate}
\item 403 instances from \bench{dfremont-project}
  (\href{https://github.com/dfremont/counting-
    benchmarks/tree/master/benchmarks/projection/application}{\nolinkurl{github:dfremont/counting-benchmarks}}),
  which is a collection of various instances originating in multiple
  domains~\cite{Fremont20a} from which we took the projection
  instances only;
\item 162 instances from \bench{2-QBF}
  (\href{https://zenodo.org/record/4292168}{\nolinkurl{Zenodo:4292168}}),
  which originate in the QBF Evaluation 2017 and
  2018~\cite{PulinaSeidl17a,PulinaSeidl18a}; and have been converted
  to a model counting problem
  (\href{https://github.com/berkeley-abc/abc}{\nolinkurl{github:berkeley-abc/abc}}~\cite{MishchenkoEtAl20a})
  contributed by Golia and Meel~\cite{FichteHecherHamiti20}.
\item 420 instances from \bench{Neural}
  (\href{https://zenodo.org/record/4292168}{\nolinkurl{Zenodo:4292168}}),
  which are instances that originate in verifying neural networks
  (\href{https://github.com/teobaluta/}{\nolinkurl{github:teobaluta}})
  contributed by Baluta, Shen, Shinde, Meel, and
  Saxena~\cite{FichteHecherHamiti20}.
\end{enumerate}

Again, we ran existing solvers on all instances with a timeout of 2
hours, namely,
\href{https://github.com/potassco/clasp}{\solver{clasp}}~\cite{GebserKaufmannSchaub12a},
\href{https://github.com/meelgroup/ganak}{\solver{Ganak}}~\cite{SharmaEtAl19a},
and
\href{http://www.cril.univ-artois.fr/KC/projmc.html}{\solver{projMC}}~\cite{LagniezMarquis19a}.
However, we had to exclude the results from projMC, since it
segfaulted notably often on our systems.
From the instances, we chose 200 instances by sampling uniformly at
random among the distribution given in
Figure~\ref{fig:mc_runtime_distribution}.
Then, we followed the same approach as on the two other tracks.
Both public and private instances, including a mapping to the original
source, are available for download at
\href{https://zenodo.org/record/3934427}{\nolinkurl{Zenodo:3934427}}~\cite{FichteHecher20a}.

\section{Competition Settings}
In the following, we state the submission requirements for the 1st
Model Counting Competition and basic information on the system on
which we ran the challenge.

\paragraph{Submission Requirements and Limits}
In order to facilitate participation of many teams, we had a very
relaxed submission policy, namely the software needs has to be
executable on the evaluation system (Linux) and initial submissions
have to be done on the cloud evaluation platform
optil.io~\cite{WasikAntczakBadura16a}.
Since our evaluation resources were limited and we were interested in
the solving behavior on a larger number of instance while allowing the
participants to have a ``training'' phase on public instances, we
restricted the runtime to 1,800 seconds (Track~1+2) and 3,600 seconds
(Track~3) and the available main memory to 8GB per instance.
Note that in general, we considered a solver to be better than
another, if it solves more instances faster than the other solver.

\paragraph{Hardware}
Finally, we evaluated the solvers on a cluster running on
Ubuntu~16.04.1 LTS Linux machines and a Linux kernel~4.4.0-184.
The cluster comprised 9 nodes each equipped with two Intel Xeon
E5-2650 CPUs consisting of 12 physical cores 256 GB RAM. We forced
performance governors to 2.2 GHz clock speed~\cite{SchoneEtAl19a},
disabled hyper threading, and enforced the process that handles the
solver invocation to run on cores 0,1,14,15 and enforced solvers to
run on cores 2--6, 7--11, 14--18, and 19--23.
We explicitly disabled transparent huge
pages~\cite{FichteMantheySchidler20a}.

\section{Participants and Results}
For the model counting competition, we received submissions from 9
teams participants coming from 8 countries and four regions: France,
Germany, India, Japan, Singapore, Poland,
USA. %
17 versions were submitted for Track 1, 11 versions for Track 2, and 6
versions for Track 3.

\subsection{Track 1: Model Counting}

\begin{figure}[t!]
  \begin{subfigure}{1\textwidth}
    \centering
    \resizebox{.8\textwidth}{!}{%
      \includegraphics{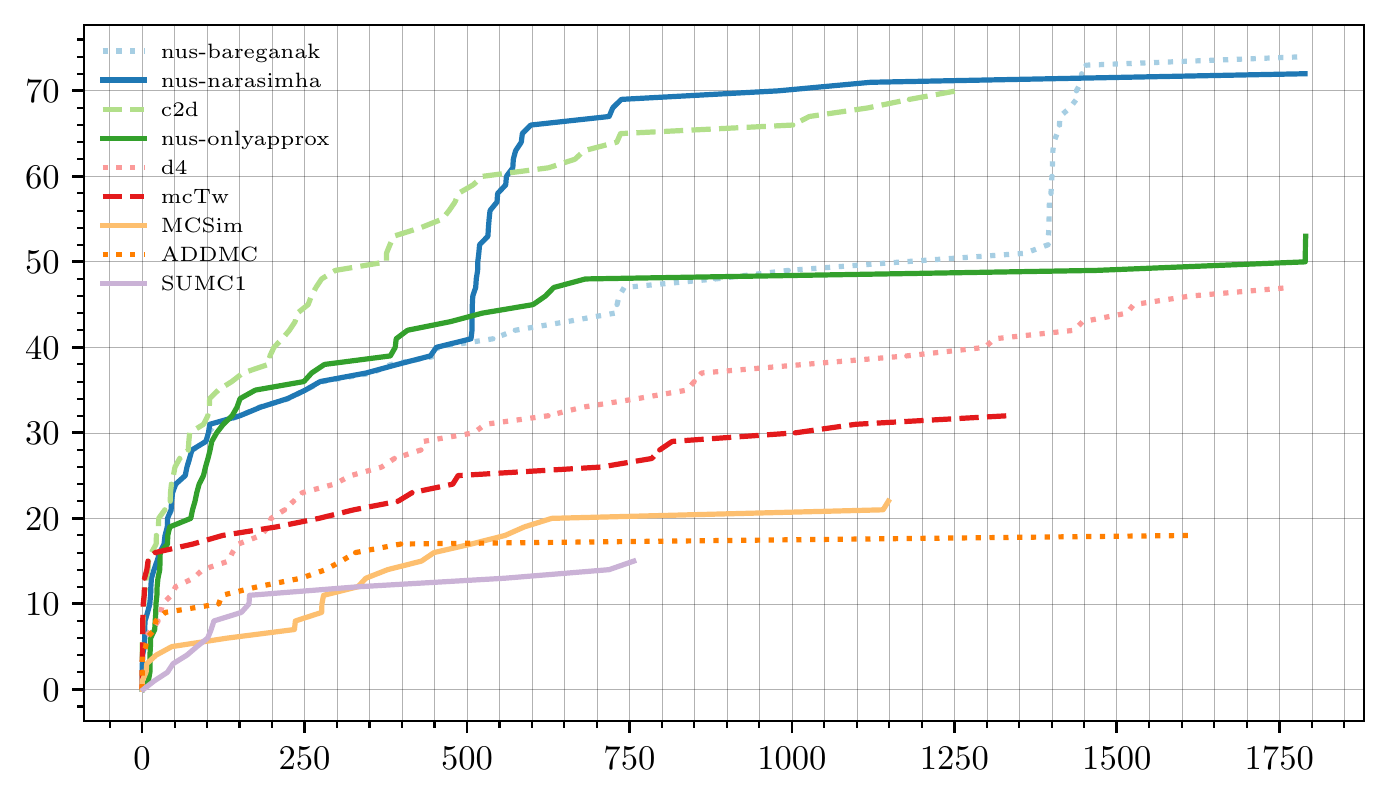}
    }
    \caption{Runtime results illustrated as cumulated solved instances. %
      The y-axis labels consecutive integers that identify instances. The
      x-axis depicts the runtime. The instances are ordered by running
      time, individually for each solver.}
    \label{fig:mc_postcactus}
  \end{subfigure}\\[2em]
  \begin{subfigure}{1\textwidth}
    \centering
    \begin{tabular}{llrrrrrHHrHrHHHH}
      \toprule
      POS &          submission &  \# & $\#_1$ & $\#_0$ & n & TLE & MEM & RTE & $t_\text{avg}[s]$ & $t_{\min}[s]$ &   $t_\text{sum}[h]$ & $t_\text{std}[s]$ & $t_\text{max}[s]$ & Download & Reference \\
      \midrule
      1 &       \solver{nus-bareganak} &  \textbf{75} &     72 &     53 & 76 & 0  & 1 & 23 & 604 &    0 & 12.8 & 617 & 1791 & TODO & TODO\\
      2 &   \solver{nus-narasimha} &  \textbf{73} &     69 &     48 & 74 & 0  & 0 & 26 & 324 &    0 & 6.7 & 326 & 1790 & TODO & TODO \\
      3 &             \solver{c2d} &  \textbf{71} &     71 &     67 & 72 & 24 & 1 & 3 & 265 &    0 & 5.3 & 320 & 1253 & TODO & TODO \\
      4 &  \solver{nus-onlyapprox}     &  \textbf{54} &     54 &     54 & 54 & 0  & 3 & 43 & 310 &    2 & 4.6 & 494 & 1791 & TODO & TODO \\
      5 &              \solver{d4} &  \textbf{48} &     48 &     47 & 48 & 43 & 1 & 8 & 501 &    0 & 6.7 & 538 & 1766 & TODO & TODO \\
      6 &           \solver{mcTw} &  \textbf{33} &     33 &     32 & 33 & 53 & 14 & 0 & 284 &    0 &  2.6 & 386 & 1332 & TODO & TODO \\
      7 &           \solver{MCSim} &  \textbf{23} &     23 &     23 & 24 & 27 & 49 & 0 & 347 &    0 &  2.3 & 313 & 1148 & TODO & na \\
      8 &           \solver{ADDMC} &  \textbf{19} &     19 &      0 & 19 & 20 & 51 & 10 & 194 &    0 &  1.0 & 369 & 1613 & 
\href{https://github.com/vardigroup/ADDMC}{\nolinkurl{H:vardigroup/ADDMC}} &  \cite{DudekPhanVardi20}\\
      9 &         \solver{SUMC1} &  \textbf{16} &     16 &     16 & 16 & 0  & 84 & 0 & 214 &    2 &  1.0 & 246 &  757 & TODO & na \\
      \bottomrule
    \end{tabular}
  \caption{Detailed standings of the submitted solvers. %
    POS refers to the position of the solver. %
    $n$ indicates the number of instances on which the solver terminated successfully. %
    \# indicates the number of  instances that have been solved with a result that was 
    within an accuracy of 10\% to the precomputed model count, %
    $\#_1$ within an accuracy of 1\%, and %
    $\#_0$ exactly as the precomputed solution. %
    TLE refers to the number of instances were the runtime limit was exceeded. 
    Note that if the sum over columns $n$ and TLE is not 100, we either observed  a memory overflow or the solver 
    terminated early without outputting a solution on the remaining instances. 
    $t_\text{avg}[s]$ contains the average runtime over all solved instances in seconds, %
    $t_\text{sum}[h]$ states the cumulative runtime over all all solved instances in hours. %
  }
  \label{tab:mc_ranking}
  \end{subfigure}
  \caption{Overview on the results for Track 1 (Model Counting) on the 100 private instances.}
\end{figure}

Figure~\ref{fig:mc_postcactus} illustrates runtime results for all
submitted solvers as CDF plot. Table~\ref{tab:mc_ranking} gives a
detailed overview on the standings and solvers. We allowed each solver
30 minutes per instance.
We ranked the solvers based on the number of solved instances for
which a model count was outputted and the model count was within a
10\% accuracy. More precisely, we precomputed the model
count~$c_\text{pre}$ for most of the instances. For instances where we
knew a model count, we marked an instance as solved accurately by a
solver if the model count~$c_\text{solver}$ outputted by the solver
satisfied the following equation:
$c_\text{pre}-\frac{1}{10}c_\text{pre} \leq c_\text{solver} \leq
c_\text{pre}+\frac{1}{10}c_\text{pre}$.
In order to open up the competition for solvers that allow only
approximate model counting, we decided not to disqualify solvers that
output solutions that are outside the accuracy interval.
We counted solutions to instances for which we did not know the model
count as successful, if the solver was the only solver that outputted
the model count and if an exact solver also outputted a solution if
the model count was within the accuracy interval.
Table~\ref{tab:mc_ranking} also contains an overview on the total
number of solutions each solver outputted as well as the number of
solved instances within accuracy 1\% (column $\#_1$) and 0\% (column
$\#_0$).

\paragraph{Winning Team}
Mate Soos, Shubham Sharma, Subhajit Roy, and Kuldeep S. Meel won this year’s Track 1
with their submission \solver{nus-barganak} by solving 75 private
instances in overall 12.8 hours at an average of 604 seconds when
considering the solved instances.
According to their submission script, the authors employ a combination
of tools, which one might consider already as a portfolio
solver. Initially, they compute the independent support using
B+E~\cite{LagniezLoncaMarquis16a} and rewrite the input instance
including the support variables for projection. Then, they run a
competition version of Ganak
(ganak\_plus\_panini)~\cite{SharmaEtAl19a}. If Ganak fails, they run
approxmc~\cite{SoosGochtMeel20,SoosMeel19}.

\paragraph{Runner-up}
Mate Soos, Shubham Sharma, Subhajit Roy, and Kuldeep S. Meel scored also the second
rank with their submission \solver{nus-narasimha} by solving 73
private instances in 6.7 hours at an average runtime of 324 seconds
over the solved instances.
According to their submission script and the shasums on the submitted
binaries, the version almost identical to the winning submission.
However, here they use fixed timeouts for Ganak~\cite{SharmaEtAl19a}
and run approxmc~\cite{SoosGochtMeel20,SoosMeel19} much earlier, which
is clearly visible from the runtime results illustrated in
Figure~\ref{fig:mc_postcactus}.

\paragraph{Third Place}
Adnan Darwiche and Arthur Choi accomplished a safe third place with
their submission \solver{c2d}. The result was in fact very close to
the team that obtained both the first and the second place on the
private instances.
If we take a look at the instances and assume very high precision, the
solver \solver{c2d} would rank even better.
While the developers experimented with
B+E~\cite{LagniezLoncaMarquis16a} as preprocessor, their final
submission ran only an updated version of the solver c2d with the
options~\ops{-in\_memory} \ops{-count} \ops{-dt\_method 6}.

\paragraph{Inconsistencies in the Outputted Model Count}
On the model counting track, we observed on that some exact solvers
outputted a solution that slightly varied from the precomputed model
count and that results were not necessarily consistent  results. Namely, on the private
instances we observed the following picture: The solver~\solver{c2d}
differed on 3 instances to the solution, which we initially
precomputed with one solver, and \solver{c2d} gave a solution that was
even outside the 10\% accuracy margin on one instance for a
precomputed solution. The submission \solver{MCSim}, which is based on
the exact solver \solver{sharpSAT}, outputted one instance a model
count that varied from the precomputed value. Similar, the solver
\solver{d4} gave on one instance a slightly different solution.

\subsection{Track 2: Weighted Model Counting}

Figure~\ref{fig:wmc_postcactus} illustrates runtime results for all
submitted solvers as CDF plot. Table~\ref{tab:wmc_ranking} gives a
detailed overview on the standings and solvers. 
We counted solutions (accuracy) and ranked the solvers as on the
previous track.
While Table~\ref{tab:wmc_ranking} also contains an overview on more
details in terms of accuracy, we refer to a more detailed discussion
below.

\newcommand{\aware}[1]{\textit{#1}\ensuremath{^*}}

\begin{figure}[t!]
  \begin{subfigure}{1\textwidth}
    \centering
    \resizebox{.8\textwidth}{!}{%
      \includegraphics{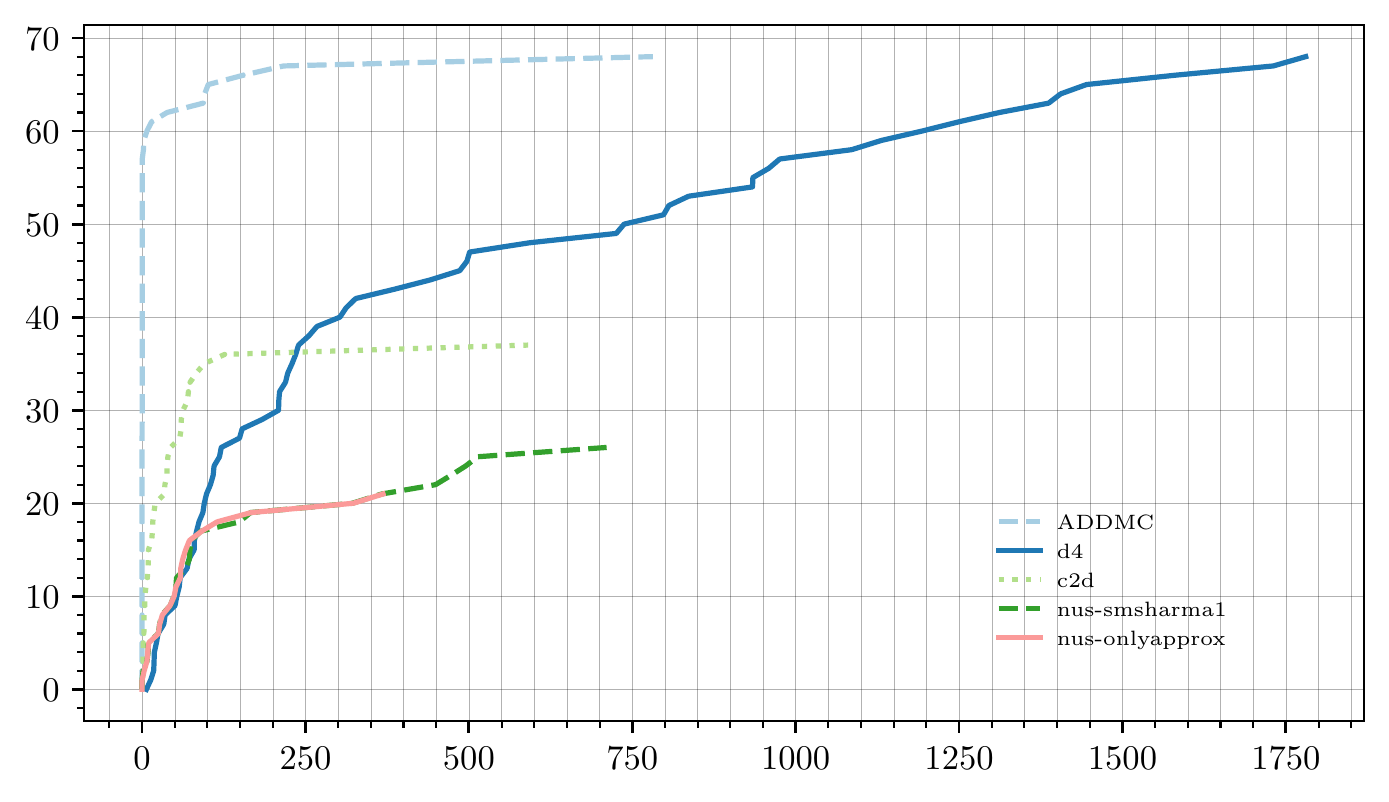}
    }%
    \caption{Runtime results illustrated as cumulated solved instances. %
      The y-axis labels consecutive integers that identify instances. The
      x-axis depicts the runtime. The instances are ordered by running
      time, individually for each solver.
    }%
    \label{fig:wmc_postcactus}
  \end{subfigure}\\[2em]
  \begin{subfigure}{1\textwidth}  
  \centering
  \begin{tabular}{llrrrrrHHrHrHHHH}
    \toprule
    POS &          submission &  \# & $\#_1$ & $\#_0$ & n & TLE & MEM & RTE & $t_\text{avg}[s]$ & $t_{\min}[s]$ &   $t_\text{sum}[h]$ & $t_\text{std}[s]$ & $t_\text{max}[s]$ & Download & Reference \\
    \midrule
    1 &              \solver{d4} &  \textbf{69} &     \aware{63} &     \aware{58} &  73 & 1 & 26 & 0 & 437 &    0 & 8.7 & 489 & 1780 & TODO & TODO \\
        &           \solver{ADDMC} &  \textbf{69} &     \aware{62} &     \aware{29} &  74 & 0 & 20 & 0 & 21  &    0 & 0.4 &  97 &  782 & TODO & TODO \\
    3 &             \solver{c2d} &  \textbf{38} &     \aware{34} &      \aware{4} &  38 & 9 & 53 & 0 & 46  &    0 & 0.5 &  96 &  591 & TODO & TODO  \\
    4 &   \solver{nus-smsharma1} &  \textbf{27} &     \aware{27} &     \aware{22} &  44 & 36 &20 & 0 & 156 &    0 & 1.9 & 190 &  711 & TODO & TODO  \\
    5 &  \solver{nus-onlyapprox} &  \textbf{22} &     \aware{22} &     \aware{17} &   94 &    38 & 0 & 62 & 0 & 0 &  1.0 & 105 &  391 & TODO & TODO \\
    \bottomrule
  \end{tabular}
  \caption{Detailed standings of the submitted solvers. %
    POS refers to the position of the solver. %
    $n$ indicates the number of instances on which the solver terminated successfully. %
    \# indicates the number of  instances that have been solved with a result that was 
    within an accuracy of 10\% to the precomputed model count, %
    $\#_1$ within an accuracy of 1\%, and %
    $\#_0$ exactly as the precomputed solution. %
    The symbol~\aware{$\cdot$} indicates that the result is unreliable due to imprecise pre-computations. 
    See discussion in the section on weighted model counting.
    TLE refers to the number of instances were the runtime limit was exceeded. 
    Note that if the sum over columns $n$ and TLE is not 100, we either observed  a memory overflow or the solver 
    terminated early without outputting a solution on the remaining instances. 
    $t_\text{avg}[s]$ contains the average runtime over all solved instances in seconds, %
    $t_\text{sum}[h]$ states the cumulative runtime over all all solved instances in hours. %
  }
  \label{tab:wmc_ranking}
\end{subfigure}
\caption{Overview on the results for Track 2 (Weighted Model Counting) on the 100 private instances.}
\end{figure}

\paragraph{Winning Teams}
The winner podium on the weighted model counting track is shared by
two teams whose submissions both solved 69 instances within a 10\%
accuracy.
One team consists of Jean-Marie Lagniez and Pierre Marquis who
submitted the solver~\solver{d4}~\cite{LagniezMarquis17a} and the
second team of Jeffrey Dudek, Vu Phan, and Moshe Vardi who submitted
the solver \solver{ADDMC} as source code under MIT
license~\cite{DudekPhanVardi20}.
While the solver \solver{d4} solves the instances in total in 8.7
hours and on average in 437 seconds with higher accuracy, the solver
\solver{ADDMC} outputted the model counts for the instances in total
in incredible 0.4 hours and 21 seconds on average. However, the solver
trades the fast runtime with a very high memory consumption.
The solver \solver{d4} runs first the preprocessor
B+E~\cite{LagniezLoncaMarquis16a} and subsequently a knowledge
compiler that transforms the input formula into a deterministic
decomposable negation normal form from which it reads the number of
solution.
In contrast, the solver \solver{ADDMC} employs dynamic programming and
use algebraic decision diagrams as data structure.

\paragraph{Third Place}
Adnan Darwiche and Arthur Choi accomplished a safe third place with
their submission \solver{c2d}~\cite{Darwiche02,Darwiche04a} by solving
38 instances within the expected 10\% accuracy.
The submission computes a deterministic decomposable negation normal
form from the input instance using the solver c2d with the
options~\ops{-in\_memory} \ops{-count} \ops{-dt\_method 6}. Then, to
obtain high accuracy, it determines the model count independently
using a small Python program.
While Table~\ref{tab:wmc_ranking} suggests that \solver{c2d} is not an
exact solver, it is in fact the most accurate solver we received. We
refer to a short discussion on accuracy below.

\paragraph{Accuracy}
In the result tables, we illustrate for each track the accuracy with
respect to the precomputed solution,~i.e., the number of instances
solved within accuracy of 1\% and 0\% of the precomputed solution.
However, we precomputed the weighted model count with the solvers
\solver{Cachet}, \solver{d4}, and \solver{miniC2D}, which output the
weighted model count only with a small number of decimal places.
Hence, if the solution is very close to 0, one obtains a high
inaccuracy to the precomputed value. 
In consequence, we need to be aware that the results presented in
Table~\ref{tab:wmc_ranking} need detailed interpretation. When looking
at the number of solved instances with higher accuracy for the solver
\solver{c2d} is seems quite inaccurate. However, the opposite is
actually the case. The approach used in \solver{c2d} provides much
higher precision than our precomputed result.
At this point we would like to thank Arthur Choi who pointed out the
following remarks: Most knowledge compilation based model counters can
save their circuits to a file and the model count can then be computed
independently with a short Python script\footnote{For (weighted) model
  counting with an arbitrarily high precision, one might require to
  use an additional library such as mpmath or gmp.}. This can be used
to ensure a high/infinite precision integer/float arithmetic.
The solver \solver{c2d} represents floating point numbers using
rational numbers, which provides a high precision for weighted model
counting.
In the probabilistic inference competitions at UAI, solvers are
required to report the log of the probability, which is the log of the
weighted model count. In fact, that value is much less likely to
underflow and a log-sum-exp trick is used to carry out the actual
arithmetic of counting.
We will likely pick up this suggestion for the next iteration.

\subsection{Track 3: Projected Model Counting}

\begin{figure}[t!]
  \begin{subfigure}{1\textwidth}
    \centering
    \resizebox{.9\textwidth}{!}{%
      \includegraphics{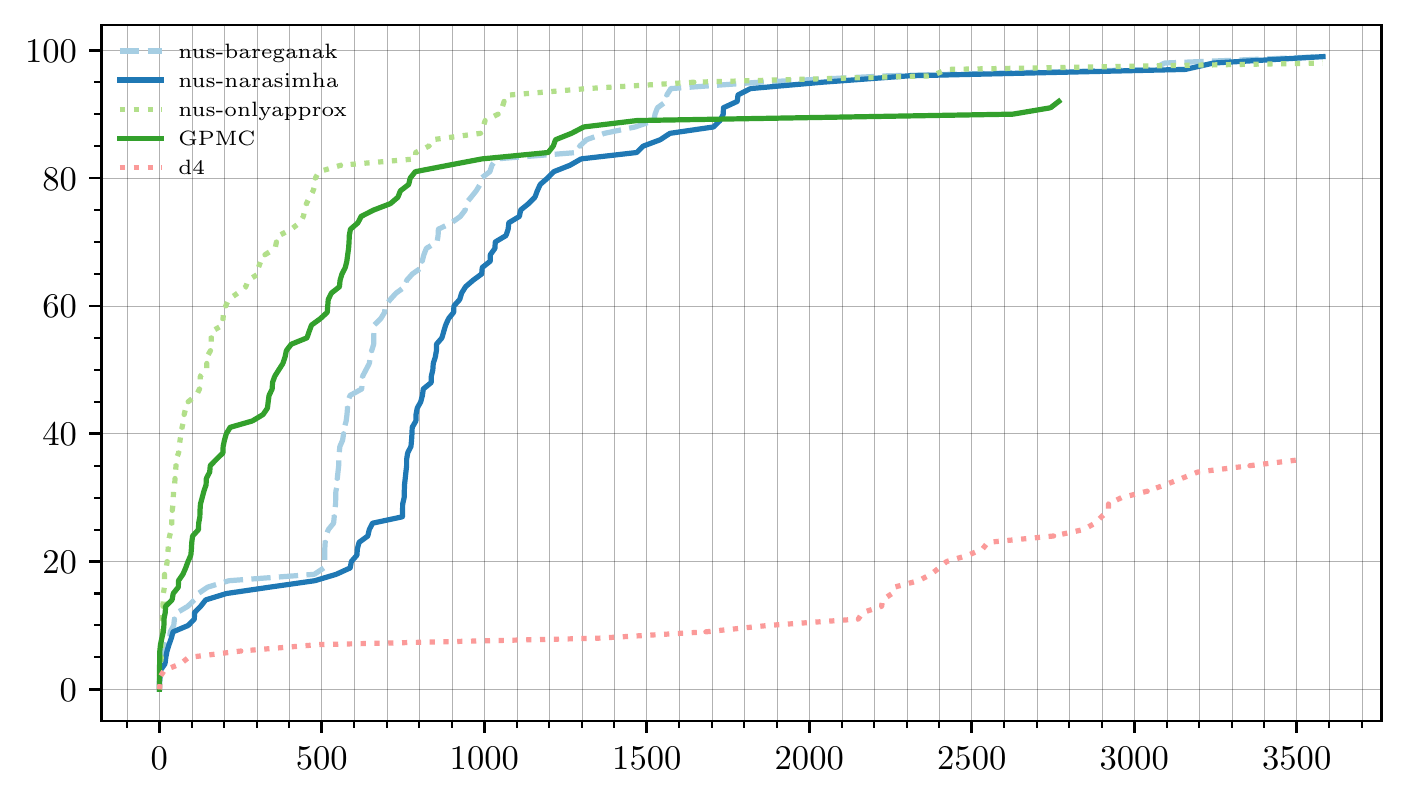}
    }%
    \caption{Runtime results illustrated as cumulated solved instances. %
      The y-axis labels consecutive integers that identify instances. The
      x-axis depicts the runtime. The instances are ordered by running
      time, individually for each solver.
    }
    \label{fig:pmc_postcactus}
  \end{subfigure}\\[2em]
  \begin{subfigure}{1\textwidth}  
  \centering
  \begin{tabular}{llrrHHHHrHrHHHH}
    \toprule
    POS & submission & \# & $\#_1$ & $\#_0$ & TLE & MEM & RTE & $t_\text{avg}[s]$ & $t_{\min}[s]$ &   $t_\text{sum}[h]$ & $t_\text{std}[s]$ & $t_\text{max}[s]$ & Download & Reference \\
    \midrule
    1 & \solver{nus-bareganak}	 & \textbf{100} & \aware{65} & 0 & ... & .... & ..... & 759 & .. & 21.1 & . & . & TODO & TODO \\ 
        & \solver{nus-narasimha}	 & \textbf{100} & \aware{100} & 0 & ... & .... & ..... & 923 & .. & 25.6 & . & . & TODO & TODO \\
    2   & \solver{nus-onlyapprox} & \textbf{99} & \aware{58} & 0 & ... & .... & ..... & 344 & .. & 9.5 & . & . & TODO & TODO \\
    3	& \solver{GPMC}	 & \textbf{93} & \aware{93} & 0 & ... & .... & ..... & 452 & .. & 11.7 & . & . & TODO & TODO \\
    4	& \solver{d4}	& \textbf{37} & \aware{37} & 0 & ... & .... & ..... & 2047 & .. & 21.0 & . & . & TODO & TODO \\
    \bottomrule
  \end{tabular}
  \caption{Detailed standings of the submitted solvers. %
    POS refers to the position of the solver. %
    $n$ indicates the number of instances on which the solver terminated successfully. %
    \# indicates the number of  instances that have been solved with a result that was 
    within an accuracy of 10\% to the precomputed projected model count and %
    $\#_1$ within an accuracy of 1\% to the  precomputed count. %
    The submission on position * is technically a version of the
    previous two submissions using only approxmc. 
    TLE refers to the number of instances were the runtime limit was exceeded. 
    Note that if the sum over columns $n$ and TLE is not 100, we either observed  a memory overflow or the solver 
    terminated early without outputting a solution on the remaining instances. 
    $t_\text{avg}[s]$ contains the average runtime over all solved instances in seconds, %
    $t_\text{sum}[h]$ states the cumulative runtime over all all solved instances in hours. %
  }
\end{subfigure}
\caption{Overview on the results for Track 3 (Projected Model Counting) on the 100 private instances.}
 \label{tab:pmc_ranking}
\end{figure}

Figure~\ref{fig:pmc_postcactus} provides and overview on runtime
results for all submitted solvers as CDF
plot. Table~\ref{tab:wmc_ranking} lists the of participating solvers,
including average runtime and total runtime of the solvers. 
Similar to the previous setting we check for 10\% range within our
precomputed value, but we allowed each solver 60 minutes per instance.

\paragraph{Winning Teams}
Mate Soos, Shubham Sharma, Subhajit Roy, and Kuldeep S. Meel from the National
University of Singapore (NUS) obtained the first place by solving 100
instances with their two submission \solver{nus-bareganak} and
\solver{nus-narasimha}, which are contrary to their names portfolio
solvers.
The submission \solver{nus-bareganak} used the solver
ganak~\cite{SharmaEtAl19a} running for 500s and if ganak did not
finish in time approxmc for the remaining time with parameter
$\epsilon=0.3$ and $\delta=0.15$~\cite{ChakrabortyEtAl14a}.
The submission \solver{nus-narasimha} runs a preprocessing step, then
ganak, but prefers approxmc~\cite{ChakrabortyEtAl14a} in the overall
runtime.
While the submission \solver{nus-bareganak} solves the instances in
total in 21.1 hours and on average in 759 seconds, the submission
\solver{nus-narasimha} outputted the projected model counts for the
instances in total in 25.6 hours and 923 seconds on average.

\paragraph{Runner-up}
The submission \solver{nus-onlyapprox} by Mate Soos, Shubham Sharma,
Subhajit Roy, and Kuldeep S. Meel ranks second. It obtained the
projected model count to only one instance less that the other two
portfolio submissions by using only approxmc~\cite{ChakrabortyEtAl14a}
instead of a portfolio.
To our surprise, its total runtime was only 9.5 hours and 344 seconds
on average.

\paragraph{Third Place}
The third place goes to Kenji Hashimoto from the Nagoya University
missing the first two ranks just by very few solved instances.
Kenji's solver \solver{GPMC} is a component caching-based solver
that was implemented on top of sharpSAT and glucose. 
It solved overall 93 in 11.7 hours at an average runtime of 454
seconds.

\paragraph{Fourth Place}
Since the NUS group submitted two portfolio solvers in variations, we
would like to point out a notable result by Jean-Marie Lagniez and
Pierre Marquis who submitted the solver \solver{d4}, which solved 37
instances at a total runtime of 21.0 hours with an average runtime of
2047 seconds.
The submission runs first a preprocessor (vivification, literal
elimination, implied literal identificaion) followed by the knowledge
compiler \solver{d4} in projected counting mode.

\subsection{General Remarks}
Overall, we were happy that many solvers performed quite well on the
challenging benchmarks. For the projected counting benchmarks, we
noticed that they might have been too easy.
In the light of knowledge compilers, it is surprising that on our
benchmark sets on Track~1 and Track~2 the results of the KC-based
solvers \solver{c2d} and \solver{d4} are flipped while using similar
approaches in principle. This suggests that they implement different
strategies, simplifications, or might even have differences in the
actual compilation approach. We think it can be interesting to peruse
why it is actually the case.
When taking a look on the rising techniques of approximate solvers, we
received interesting, fast, and quite accurate submissions, however,
it is worth pointing out that the submissions seem to profit a lot
from a solving portfolio that also includes exact solving techniques.
A great-performing newcomer in projected model counting was the
solver \solver{GPMC}, which is based on component-caching techniques.
During the competition we ran into a few organizational and technical
issues, which we quickly want to discuss below.
\paragraph{Accuracy, Correctness, Stability, and Verification of Results}
In contrast to what we heard from early SAT competitions, we noticed
that most of the submitted solvers are very stable and produced no
segfaults or entirely unreliable results.
Still, accuracy and correctness of the various solvers might be
improved.
Above we mentioned that judging on the accuracy and correctness of the
different solvers should be improved for the next iteration. So far,
we precomputed a model count, but stored the result with lower
precision for the weighted model counting instances. In consequence,
an exact solver with much higher precision shows on quite a number of
instances lower accuracy with respect to our na\"ive metric whereas it
is actually much more precise. This problem might even result in
instances shown as inaccurately solved when the numbers get closer to
0. Hence, we suggest for a next iteration to use as output the log of
the weighted model count.
Further, we observed cases where two solvers outputted different
results, a situation that we also saw with exact solvers.
Hence, we are facing elaborate questions that might originating in (i)
bad accuracy, more specifically, (ia) having a method that is not
exact (which is a conceptual topic of approximation); (ib) having
inappropriate arithmetic precision such as fixed-point numbers with
exact methods (which might happen in component-caching based solvers
or DP-based solvers); (ic) having an output at low precision while the
result was computed correctly (which might easily happen with
knowledge compilers); or (ii) bugs in a supposedly exact solver.
From our perspective, this clearly suggests new developments and
further research.
For Issue~(ia) one could only allow for exact techniques or explicitly
mark imprecise techniques. For Issues~(ib) and (ic), a different way of
outputting the information or novel output format might help.
For addressing the correctness as stated in Issue~(ii), we believe
that it will be useful to design techniques to prove the exact model
count for model counting in order to validate the actual model
count. Since various model counters are based on CDCL-solving,
extending techniques similar to propositional
satisfiability~\cite{WetzlerHeuleHunt14a} and ideas in knowledge
compilation~\cite{Capelli19} or probabilistic
inference~\cite{KarimiKaskiKoivisto20} might a direction to pursue.

\paragraph{Execution}
In the beginning, we used the platform Optil.io, which provided us
with a uniform interface when handling submissions where we could on
top see a leader board during an active competition phase.
The idea with the leader board did not pan out as we could
open the submissions not much in advance and then some groups
submitted at the very last moment.
On top, some contestants were mostly used to the StarExec system,
required to submit dynamically linked binaries, and needed to generate
temporary output.
Since the submission system was not primarily designed for this use,
debugging of the submissions was seen in some cases as quite
cumbersome.
To provide a uniform and reproducible evaluation platform, we decided
to run the experiments on the Taurus Cluster~\cite{taurus}, which we
could unfortunately not complete due to a major security incident on
European high performance research clusters~\cite{Cimpanu20}.
Luckily, Stefan Woltran and Toni Pisjak helped out on very short
notice by providing resources on a cluster at TU Wien.
Still, we went far beyond our initially intended schedule and finished
some results only a day prior to the presentation at the SAT
conference.
In the competition, we used the tool runsolver~\cite{Roussel11} to
control the execution. It is known that this tool suffers from the
sampling based issues, namely measuring resources results in
immediately expired information and enforcing limits might not do
anything if the used RSS (resident set size) exceeds the indented
maximum limit only in sudden resource
spikes~\cite{BeyerLoweWendler15,BeyerLoweWendler19}.
We did not observe immediate indicators that would prevent using
runsolver in our setting. Still, we suggest to use a cgroups based
system in the future, while knowing that the system does not work out
of the box and installing this system requires root privileges.
For the next iteration, we suggest to try StarExec and Mate Soos
offered support in configurating the system.

\paragraph{Data Format}
We suggested a data format intending to have a distinguishable format
for the tracks while keeping the format very close to the original
DIMACS format and removing certain ambiguities. Since a group
complained as it required reencoding the headers and breaking
downwards compatibility, which might confuse users of the solvers. We
will likely suggest an update to the headers and weights on \WMC for
the next iteration.

\paragraph{Housekeeping}
While the solvers performed very well and appeared quite stable in
terms of the produced results, some groups followed a laissez-faire
approach when constructing their submission scripts.
Some solvers did not clean up after solving leaving gigabytes of
temporary data on the disk assuming that temporary directories are
handled by the user or would just remove everything starting with a
certain pattern in a directory when starting the script.
Others, did not implement proper signal handling or included a fixed
timeout.
While this is clearly reasonable in a competition setting, we do not
see that it fosters practical applicability of the submissions.
Therefore, we suggest the definition of uniform exit codes/return
values for the next iteration to avoid full log file parsing when
evaluating the results and allow for simplifying the evaluation
scripts.
Further, we suggest not to announce a hard timeout, but instead to use
a vague cutoff interval. This enforces to implement clean handling of
unix system signals and not to optimize and hardcode fixed timeouts.

\paragraph{Judge}
During the competition, we ran into a situation where we were
imprecise with the initial competition specifications, made mistakes
during the execution, or evaluation.
So far, we were hopefully able to sort out all issues and
complaints. However, for the next iteration, we suggest to setup one
or two persons who serve as judge(s). The idea is that a judge makes
final decisions if the organizers and a submitter disagree on a rule
or one has to add amendments to a rule if an unforeseen situation occurs.

\section{Organization}

The composition of the program committee during MC 2020 was as
follows:

\begin{table}[h]
  \centering
  \begin{tabular}{lll}
    \textbf{Program Committee}
    &Johannes Fichte & TU Dresden\\
    &Markus Hecher & TU Vienna \& University of Potsdam\\
    \textbf{Student Assistant}
    &Florim Hamiti & TU Dresden\\
  \end{tabular}
\end{table}

\section{Conclusion and Future}

We thank all the participants for their enthusiasm and strong and
interesting contributions. Special thanks go to the participants who
also presented work at the Model Counting Workshop at the SAT 2020
conference. We are very happy that this edition attracted many groups. 
While we initially ran into a few hiccups during the submission and
execution phase, we were happy about strong contributions and hope
that the initial execution will lead to more future editions spawning
new application directions of model counting.

For the competition, we did not enforce strong requirements for
submissions by allowing any kind of external libraries, binaries and
by not asking for open source or public repositories of the submissions.
After the competition, we released the benchmark instances and we
included na\"ive statistics about the instances in our report.
In future, it might be interesting to investigate whether techniques
for analyzing instances such as community structures in SAT solving
are also interesting in the context of model
counting~\cite{AnsoteguiGiraldez-CruLevy12}.  
We do not believe that the selected competition instances necessarily
provide a good picture for future solvers. The competition is really
just a snapshot on the current state. There might be solvers that
perform well on a specific set of instances or application, then
showing still practical while performing bad on this years
selection. Therefore, we also released the full set of instances,
which we collected or received.
While this years instance selection process was fairly ad-hoc, a more
sophisticated approach might prove
helpful~\cite{HoosKaufmannSchaub13a}.
For future editions, we think that detailed solver descriptions could
be interesting.

We welcome anyone who is interested in the competition to send us an
email for receiving updates and joining the discussion on Slack. We
look forward to the next edition. Detailed information will be posted
on the website at modelcounting.org.

\section*{Acknowledgements}
We would like to thank the following people for helping with the
organization and suggestion for future editions.
Jan Badura from Optil.io for supporting us during the initial phase
with a platform for submitting solvers~\cite{WasikAntczakBadura16a}.
Adnan Darwiche and Arthur Choi for pointing us to the evaluation
procedures in the UAI competition for accuracy in solving and for
their comments on accuracy/correctness of solvers and a new output
format.
Mate Soos for detailed comments on properties of the instances and
visualizing the results.
Stefan Woltran and Toni Pisjak for providing and freeing up Cluster
resources at TU Wien on very short notice.

\appendix

\pagebreak

\newpage
\section{Short Solver Descriptions}
\paragraph{\solver{ADDMC}}
is a solver that computes the exact literal-weighted model counts of
CNF formulas. The algorithm employs dynamic programming and uses
Algebraic Decision Diagrams as the main data structure~\cite{DudekPhanVardi20}.

\paragraph{\solver{c2d}}
is a compiler that converts CNF into d-DNNF circuits, on which model
counting and weighted model counting can be performed in time linear
in the circuit size (hence, c2d is an exact MC/WMC
solver)~\cite{Darwiche02,Darwiche04a}.

\paragraph{\solver{d4}}
d4 is a compiler associating with an input CNF formula an equivalent
representation from the language Decision-DNNF. Decision-DNNF is the
language consisting of the Boolean circuits with a single output (its
root), where each input is a literal or a Boolean constant, and each
internal gate is either a decomposable $\wedge$ gate of the form
$N = \wedge(N_1, . . . , N_k)$ (``decomposable'' means here that for
each $i$, $j \in \{1,\ldots, k\}$ with $i \neq j$ the subcircuits of
$N$ rooted at $N_i$ and $N_j$ do not share any common variable) or
decision gates of the form $N = \text{ite}(x, N_1, N_2$). $x$ is the
decision variable at gate $N$, it does not occur in the subcircuits
$N_1$, $N_2$, and ite is a ternary connective whose semantics is given
by $\text{ite}(X, Y, Z) = (\neg X \wedge Y ) \vee (X \wedge Z)$
(``ite'' means ``if ... then ... else ...: if X then Z else
Y'')~\cite{LagniezMarquis17a}.

\paragraph{\solver{GPMC}}
is an exact projected model counter, which computes the number of
models of a given formula that are different when models are
restricted to the projected variables. It is a model counter combined
clause learning with component decomposition and component
caching. The underlying idea to deal with the projected model counting
is the same as DPLL-based model counters that restrict search to
projection variables and use a SAT solver for components with no
unassigned projection variables, i.e., (a) GPMC selects a branch
decision variable from unassigned projection variables first. (b) When
there is no unassigned projection variables in a component, GPMC
starts solving the satisfiability of the component. If the result is
SAT, the number of models of the component is 1; otherwise 0.

\paragraph{\solver{mcTw}}
implements in the core an algorithm that is based on dynamic
programming on treewidth decompositions of a primal graph constructed
for a given CNF formula~\cite{SamerSzeider10b}.

\paragraph{nus-barganak/nus-narasimha}
are submissions that feature portfolio solvers, consisting of
B+E~\cite{LagniezLoncaMarquis16a}, Ganak~\cite{SharmaEtAl19a}, and
approxmc~\cite{SoosGochtMeel20,SoosMeel19}. The exact configuration
depends on the track. We refer to details on the medalists of each
track.

\paragraph{MCSim}
is a component caching based solver implemented on top of
\solver{sharpSAT}~\cite{Thurley06a}.

\paragraph{SUMC1} counts how many models are eliminated by each clause
because they fail to satisfy it. Then, by computing the cardinality of
a union of sets to determine how many models are eliminated overall it
obtains the overall model count without explicitly identifying the
models. The source code is available at
\href{https://www.github.com/ivor-spence/sumc}{\nolinkurl{github:ivor-spence/sumc}}~\cite{Spence20}.

\section{Data Formats}

\newcommand{\fsym}[1]{\texttt{#1}}
\subsection{Track 1: Model Counting}\label{sec:dataformat:mc}

The input format for providing a formula (\emph{.mcc2020\_cnf}) was
taken from the DIMACS-format for formulas in conjunctive normal
form~\cite{TrickChvatalCook93a}. The DIMACS-input format is used in
SAT competitions. For more details, we refer to an online resource at
\url{http://www.satcompetition.org/2009/format-benchmarks2009.html}
We use the following version where we print symbols in typewriter
font, e.g., \fsym{$\backslash{}$n}

\begin{itemize}
\item Line separator is the symbol \fsym{$\backslash{}$n}.
\item Lines starting with character \fsym{c} are interpreted as comments.
\item Variables are consecutively numbered from \fsym{1} to \fsym{n}.
\item The problem description is given by 
  a unique line of the form
  \fsym{p cnf NumVariables NumClauses} that we expect to be the first
  line (except comments). More precisely, the line starts with
  character p (no other line may start with p), followed by the
  problem descriptor \fsym{cnf}, followed by number \fsym{n} of
  variables followed by number \fsym{m} of clauses each symbol is
  separated by space each time.
\item The remaining lines indicate clauses consisting of decimal
  integers separated by space.  Lines are terminated by character
  \fsym{0}.  The Line \fsym{2 -1 3 0$\backslash{}$n} indicates the
  clause “2 or not 1 or 3”.
\item Empty lines or lines consisting of spaces may occur and only
  will be ignored.
\end{itemize}

\begin{example}
  \noindent 
  The following text describes the CNF formula (set of clauses)
  \[\{\{\neg x_1, \neg x_2\}, \{x_2, x_3, \neg x_4\}, \{x_4, x_5\},
  \{x_4, x_6\}\}\] including a problem description line and two
  comments.\newline

\begin{verbatim}
c This file describes a DIMACS CNF in MC 2020 format 
c with 6 variables and 4 clauses 
p cnf 6 4
-1 -2 0
2 3 -4 0
c this is a comment and will be ignored
4 5 0
4 6 0
\end{verbatim}

\noindent The solution is given in the following format:
\begin{verbatim}
c This file describes that the model count is 22
s mc 22
\end{verbatim}
  
\end{example}

\subsection{Track 2: Weighted Model Counting}\label{sec:dataformat:wmc}

The input format for providing a formula (\emph{.mcc2020\_wcnf}) was
taken from the DIMACS-format for formulas in conjunctive normal
form~\cite{TrickChvatalCook93a}. The DIMACS-input format is used in
SAT competitions. For more details, we refer to an online resource at
\url{http://www.satcompetition.org/2009/format-benchmarks2009.html}.
Cachet used a similar format\footnote{See: file README.txt in
  \url{https://www.cs.rochester.edu/u/kautz/Cachet/cachet-wmc-1-21.zip}.},
however, the current avoids implicit assumptions about weights.
We use the following version where we print symbols in typewriter
font, e.g., \fsym{$\backslash{}$n}

\begin{itemize}
\item Line separator is the symbol \fsym{$\backslash{}$n}.
\item Lines starting with character \fsym{c} are interpreted as comments.
\item Variables are consecutively numbered from \fsym{1} to \fsym{n}.
\item The problem description is given by 
  a unique line of the form
  \fsym{p wcnf NumVariables NumClauses} that we expect to be the first
  line (except comments). More precisely, the line starts with
  character p (no other line may start with p), followed by the
  problem descriptor \fsym{wcnf}, followed by number \fsym{n} of
  variables followed by number \fsym{m} of clauses each symbol is
  separated by space each time.
\item The weight function is given by lines of the form \fsym{w
    Literal Weight 0} defining the floating point Weight for Literal,
  where $0 \leq \text{Weight} \leq 1$. 
  We assume that no more than 9 significant digits are given after the
  decimal point.
  If the weight for a literal is not defined, it is considered to be
  of weight 1.
\item The remaining lines indicate clauses consisting of decimal
  integers separated by space.  Lines are terminated by character
  \fsym{0}.  The Line \fsym{2 -1 3 0$\backslash{}$n} indicates the
  clause “2 or not 1 or 3”.
\item Empty lines or lines consisting of spaces may occur and only
  will be ignored.
\end{itemize}

\begin{example}
  \noindent 
  The following text describes the CNF formula (set of clauses)
  \[\{\neg x_1, \neg x_2\}, \{x_2, x_3, \neg x_4\}, \{x_4, x_5\},
    \{x_4, x_6\}\}\] with weight
  function~$\{x_1 \mapsto 0.4, \neg x_1 \mapsto 0.6, x_2 \mapsto 1,
  \neg x_2 \mapsto 1, x_3 \mapsto 1, \neg x_3 \mapsto 1, x_4 \mapsto
  0.5, \neg x_4 \mapsto 0.5, x_5 \mapsto 1, \neg x_5 \mapsto 1\}$
  including a problem description line and two comments.
  \newline

\begin{verbatim}
c This file describes a weighted CNF in MC 2020 format with 6 variables and 4 clauses 
p wcnf 6 4
w 1 0.4 0
w -1 0.6 0
w 4 0.5 0
w -4 0.5 0
w 5 1.0 0
w -5 1.0 0
-1 -2 0
2 3 -4 0
c this is a comment and will be ignored
4 5 0
4 6 0
\end{verbatim}
  
\noindent The solution should be given in the following format:
\begin{verbatim}
c This file describes that the weighted model count is 6.0
s wmc 6.0
\end{verbatim}
\end{example}  

\subsection{Track 3: Projected Model Counting}\label{sec:dataformat:pmc}

The input format for providing a formula (\emph{.mcc2020\_pcnf}) was
taken from the DIMACS-format for formulas in conjunctive normal
form~\cite{TrickChvatalCook93a}. The DIMACS-input format is used in
SAT competitions. For more details, we refer to an online resource at
\url{http://www.satcompetition.org/2009/format-benchmarks2009.html}.
Ganak used a similar format\footnote{See
  \url{https://github.com/meelgroup/ganak}} stating the projection
variables as comment.
We use the following version where we print symbols in typewriter
font, e.g., \fsym{$\backslash{}$n}

\begin{itemize}
\item Line separator is the symbol \fsym{$\backslash{}$n}.
\item Lines starting with character \fsym{c} are interpreted as comments.
\item Variables are consecutively numbered from \fsym{1} to \fsym{n}.
\item The problem description is given by 
  a unique line of the form
  \fsym{p pcnf NumVariables NumClauses} that we expect to be the first
  line (except comments). More precisely, the line starts with
  character p (no other line may start with p), followed by the
  problem descriptor \fsym{wcnf}, followed by number \fsym{n} of
  variables followed by number \fsym{m} of clauses each symbol is
  separated by space each time.
\item Projection variables, i.e., variables that are important and
  which are the ones that will be considered for the count, are given
  by a line of the form~\fsym{vp VARID1 VARID2 VARID3 0}. The line is
  expected to be unique meaning that no other line may start with
  vp. The line may occur at any time after the p line, especially for
  encoding the line may also occur as last line in the file.
  VARIDX represents decimal integers which such that
  $1 \leq VARIDX \leq n$ and where $n$ refers to the number of
  variables. The integers are separated by space and the line is
  terminated by character \fsym{0}. For example, line \fsym{vp 1 2 0
    $\backslash$n} indicates the set $\{1, 2\}$.
\item The remaining lines indicate clauses consisting of decimal
  integers separated by space.  Lines are terminated by character
  \fsym{0}.  The Line \fsym{2 -1 3 0$\backslash{}$n} indicates the
  clause “2 or not 1 or 3”.
\item Empty lines or lines consisting of spaces may occur and only
  will be ignored.
\end{itemize}

\begin{example}
  \noindent 
  The following text describes the CNF formula (set of clauses)
  \[\{\neg x_1, \neg x_2\}, \{x_2, x_3, \neg x_4\}, \{x_4, x_5\},
    \{x_4, x_6\}\}\] with projection set~~$\{x_1, x_2\}$ including a
  problem description line and two comments.
  \newline

\begin{verbatim}
c This file describes a projected CNF in MC 2020 format
c with 6 variables and 4 clauses and 2 projected variables 
p pcnf 6 4 2
vp 1 2 0
-1 -2 0
2 3 -2 0
c this is a comment and will be ignored
4 5 0
4 6 0
\end{verbatim}

\noindent The solution should be given in the following format:
\begin{verbatim}
c This file describes that the projected model count is 3
s pmc 3
c  
\end{verbatim}

\end{example}
\end{document}